\documentclass[sn-mathphys,Numbered]{sn-jnl}

\usepackage{graphicx}%
\usepackage{multirow}%
\usepackage{amsmath,amssymb,amsfonts}%
\usepackage{amsthm}%
\usepackage{mathrsfs}%
\usepackage[title]{appendix}%
\usepackage{xcolor}%
\usepackage{textcomp}%
\usepackage{manyfoot}%
\usepackage{booktabs}%
\usepackage{algorithm}%
\usepackage{algorithmicx}%
\usepackage{algpseudocode}%
\usepackage{listings}%

\theoremstyle{thmstyleone}%

\theoremstyle{thmstyletwo}%

\theoremstyle{thmstylethree}%

\begin{document}

\title[Article Title]{Quantum Entanglement Theory and Its Generic Searches in High Energy Physics
}

\author[1]{\fnm{Junle} \sur{Pei}}\email{peijunle@hnas.ac.cn}

\author[2,3]{\fnm{Yaquan} \sur{Fang}}\email{fangyq@ihep.ac.cn}

\author*[4]{\fnm{Lina} \sur{Wu}}\email{wulina@xatu.edu.cn}

\author*[2,3]{\fnm{Da} \sur{Xu}}\email{xuda@ihep.ac.cn}

\author[2,3]{\fnm{Mustapha} \sur{Biyabi}}\email{bmustapha@ihep.ac.cn}

\author*[5,6,7]{\fnm{Tianjun} \sur{Li}}\email{tli@itp.ac.cn}

\affil[1]{\orgdiv{Institute of Physics}, \orgname{Henan Academy of Sciences}, \orgaddress{\city{Zhengzhou}, \postcode{450046}, \country{P. R. China}}}

\affil[2]{\orgdiv{Institute of High Energy Physics}, \orgname{Chinese Academy of Sciences}, \orgaddress{\city{Beijing}, \postcode{100049}, \country{P. R. China}}}

\affil[3]{\orgname{University of Chinese Academy of Science}, \orgaddress{\street{No. 19A Yuquan Road}, \city{Beijing}, \postcode{100049}, \country{P. R. China}}}

\affil[4]{\orgdiv{School of Sciences}, \orgname{Xi'an Technological University}, \orgaddress{\city{Xi'an}, \postcode{710021}, \country{P. R. China}}}

\affil[5]{\orgdiv{School of Physics}, \orgname{Henan Normal University}, \orgaddress{\city{Xinxiang}, \postcode{453007}, \country{P. R. China}}}

\affil[6]{\orgdiv{Institute of Theoretical Physics}, \orgname{Chinese Academy of Sciences}, \orgaddress{\city{Beijing}, \postcode{100190}, \country{P. R. China}}}

\affil[7]{\orgdiv{School of Physical Sciences}, \orgname{University of Chinese Academy of Sciences}, \orgaddress{\street{No. 19A Yuquan Road}, \city{Beijing}, \postcode{100049}, \country{P. R. China}}}

\abstract{We propose a new formalism for quantum entanglement (QE), and study its generic searches at the colliders. For a general quantum system with $N$ particles, we show that the quantum space (the total spin polarization parameter space) is complex projective space, and the classical space (the spin polarization parameter space for classical theory) is the cartesian product of the complex projective spaces. Thus, the quantum entanglement space is the difference of these two spaces. For the $ff$, $AA$, $Af$, $fff$, and $ffA$ systems, we propose their discriminants $\Delta_i$. The corresponding classical spaces are the discriminant locus $\Delta=0$  for $ff$ system, and intersections of the discriminant loci $\Delta_i=0$ for $AA$, $Af$, $fff$, and $ffA$ systems in the quantum space. 
 In particular, for two fermion $ff$ system, we prove that our discriminant criterion is equivalent to the original Peres-Horodecki criterion and the CHSH criterion. 
And thus our quantum entanglement space is indeed Bell non-local.
With the collider searches, we can reconstruct the discriminants from various measurements, and probe the quantum entanglement spaces via a fundamental approach at exact level. In addition, for the specific approach, we present a comprehensive framework to detect quantum entanglement in high-energy multi-particle systems, spanning fermion pairs ($t\bar{t}$, $\tau^{+}\tau^{-}$), bosonic pairs ($W^{-}W^{+}$), and hybrid or three-body systems ($W^{-}t$, $ttt$, $t\bar{t}W^{-}$). By leveraging phase-space integration and the orthogonality properties of Wigner $d$-functions, we derive diverse observables through angular correlations in decay products and establish key criteria for QE by calculating the exact quantum ranges and classical ranges. Analytical and numerical studies reveal that while the widely used observable $D = -3\langle\cos\theta_{e^{+}e^{-}}\rangle$ fails to detect QE in $t\bar{t}$ systems at $e^{+}e^{-}$ colliders within the Standard Model (SM), alternative observables such as $D' =\frac{32}{\pi^2} \langle\cos(\phi_{e^{+}}-\phi_{e^{-}})\rangle$ provide robust signatures of entanglement. Beyond the SM, scenarios involving Higgs-like scalars ($h'$) or heavy $b'$ quarks demonstrate enhanced entanglement through distinct angular correlations in $t\bar{t}$, $W^{-}W^{+}$, and $W^{-}t$ systems, with sensitivity to specific mass thresholds. For three-body systems like $ttt$ and $t\bar{t}W^{-}$, triple-product correlations are shown to resolve genuine tripartite entanglement. These results establish model-independent methodologies for probing QE across collider experiments, bridging quantum information principles with high-energy phenomenology, while offering novel pathways to explore exotic particles and quantum properties in multi-particle systems.
}


\maketitle

\section{Introduction} \label{sec:1}

Both quantum mechanics and special relativity are the cornerstones for modern physics. Unlike the classical physics, quantum mechanics has a couple of the most genuine features: quantum entanglement and Bell's theorem. For a quantum mechanics system, quantum entanglement is a correlation between its sub-systems, {\it i.e.}, one sub-system cannot be described independently of the other sub-system, even if they are space-like separated. In addition, Bell's theorem~\cite{PhysicsPhysiqueFizika.1.195} states that with a few basic measurement assumptions, quantum mechanics is not compatible with the local hidden variable theories. To be concrete, Bell inequalities can be satisfied by any classical theory, or in general by any local theory with hidden variables. However, they can be violated in quantum theory, which is called Bell non-locality. Quantum entanglement and Bell non-locality are closely related. However, there is a subtle difference: quantum entanglement is a necessary condition for Bell non-locality, but not a sufficient condition~\cite{Werner:1989zz}.
Historically speaking, this was first highlighted by Einstein, Podolsky, and Rosen (EPR) in 1935~\cite{PhysRev.47.777} as a challenge to the completeness of quantum mechanics. Over decades, quantum entanglement and Bell non-locality have been rigorously confirmed through various experiments violating Bell inequalities \cite{PhysicsPhysiqueFizika.1.195,PhysRevLett.28.938,PhysRevLett.47.460,PhysRevLett.49.1804,PhysRevLett.81.3563,Hensen:2015ccp} and demonstrations of quantum teleportation \cite{PhysRevLett.70.1895,Bouwmeester_1997,PhysRevLett.80.1121,Riebe:2004jpa,Barrett:2004bxh}, primarily in low-energy systems such as photons, ions, and solid-state qubits. However, the exploration of entanglement in high-energy particle physics remains an emerging frontier \cite{vonKuk:2025kbv,Subba:2024mnl,Subba:2024aut}, where the interplay between quantum correlations and relativistic dynamics opens new avenues to probe fundamental physics.  

The Standard Model (SM) for particle physics is described by quantum field theory, which is based on quantum mechanics and special relativity. Thus, we can probe the fundamental properties of quantum mechanics at various colliders. In particular, recent breakthroughs by the ATLAS and CMS Collaborations at the Large Hadron Collider (LHC) have reported evidence of quantum entanglement in top quark-antiquark ($t\bar{t}$) pairs \cite{ATLAS2024,CMS:2024pts}. The polarization state of the $t\bar{t}$ system is encoded in its spin density matrix $\rho$, which can be parameterized as  
\begin{align}  
    \rho &= \frac{I_{4} + \sum_{i}\left(B_{i}^{+}\sigma^{i} \otimes I_{2} + B_{i}^{-}I_{2} \otimes \sigma^{i}\right) + \sum_{i, j} C_{ij} \sigma^{i} \otimes \sigma^{j}}{4} \nonumber  \\
    &= {\footnotesize \frac{1}{4} 
\begin{bmatrix}
1 + B_3^+ + B_3^- + C_{33} & B_1^- + C_{31} - i(B_2^- + C_{32}) & B_1^+ + C_{13} - i(B_2^+ + C_{23}) & C_{11} - C_{22} - i(C_{12} + C_{21}) \\
B_1^- + C_{31} + i(B_2^- + C_{32}) & 1 + B_3^+ - B_3^- - C_{33} & C_{11} + C_{22} + i(C_{12} - C_{21}) & B_1^+ - C_{13} - i(B_2^+ - C_{23}) \\
B_1^+ + C_{13} + i(B_2^+ + C_{23}) & C_{11} + C_{22} + i(C_{21} - C_{12}) & 1 - B_3^+ + B_3^- - C_{33} & B_1^- - C_{31} - i(B_2^- - C_{32}) \\
C_{11} - C_{22} + i(C_{21} + C_{12}) & B_1^+ - C_{13} + i(B_2^+ - C_{23}) & B_1^- - C_{31} + i(B_2^- - C_{32}) & 1 - B_3^+ - B_3^- + C_{33}
\end{bmatrix}}~,\label{rho1}
\end{align}  
where $B_i^{\pm}$ are respectively the spin polarizations of $t$ and $\bar{t}$,
and $C_{ij}$ characterizes their spin correlations. For the decay processes $t \to e^+ + \nu_e + b$ and $\bar{t} \to e^- + \bar{\nu}_e + \bar{b}$, the angular distributions of the $e^+$ and $e^-$ in their parent particles' rest frames are governed by  \cite{Brandenburg_2002,Bernreuther:1997gs}
\begin{align}  
    \frac{1}{\sigma} \frac{d\sigma}{d\Omega_{+} d\Omega_{-}} = \frac{1 + \mathbf{B}^{+} \cdot \hat{\mathbf{q}}_{+} - \mathbf{B}^{-} \cdot \hat{\mathbf{q}}_{-} - \hat{\mathbf{q}}_{+} \cdot \mathbf{C} \cdot \hat{\mathbf{q}}_{-}}{(4\pi)^2}~,  
\end{align}  
where $\hat{\mathbf{q}}_{\pm}$ denote the momentum directions of the decay leptons. The LHC experiments utilized the observable  
$$D = \mathrm{tr}\left[\mathbf{C}\right]/3 = -3\langle \cos\theta_{e^- e^+}\rangle~,$$  
where $\theta_{e^- e^+}$ is the angle between the $e^-$ and $e^+$ directions respectively in their parent particles' rest frames,
and $\cos\theta_{e^-e^+}=\hat{\mathbf{q}}_{+}\cdot\hat{\mathbf{q}}_{-}$. A value of $D < -\frac{1}{3}$ serves as a sufficient (though not necessary) condition to certify entanglement in the $t\bar{t}$ system \cite{Afik:2020onf}. However, at $e^-e^+$ colliders, the leading-order (LO) calculations predict $D \equiv \frac{1}{3}$, independent of beam energy, polarization, or the top quarks’ emission angles. This renders $D$ ineffective for detecting entanglement at such colliders, necessitating alternative observables.  

As we know, the quantum entanglement and Bell's non-localities in the previous studies are all defined via the inequalities. While in mathematics, we perform the exact calculations. Therefore, the great challenge question is: can we propose a Quantum Entanglement Theory (QET) which can define the quantum entanglement exactly? With this theory, we can provide the solid foundation for quantum entanglement, and probe it via a fundamental approach at the exact level in general. Similarly, for any specific approach to probe the quantum entanglement, we can define the corresponding quantum entanglement criterion exactly as well.

In this paper, we propose a brand new formalism for quantum entanglement, and study its generic searches in high energy physics. Let us explain the convention first. We define the total spin polarization parameter space as quantum space, the spin polarization parameter space for classical theory as classical space, and the difference of these two spaces as quantum entanglement space. Thus, quantum space is the sum of classical space and quantum entanglement space. Also, we define the quantum non-locality tests as the tests for quantum entanglement space via the space-like separated measurements. Moreover, for a two particle system, we define the Bell local parameter space as the space which is consistent with the local hidden variable theory and 
satisfies the Clauser-Horne-Shimony-Holt (CHSH) inequality~\cite{Clauser:1969ny}, 
and the Bell non-local parameter space as the space which violate the CHSH inequality. In general,
the classical space is a subset of or equal to the Bell local parameter space, and
the Bell non-local parameter space is a subset of or equal to the quantum entanglement space. 
In addition, for a particle with spin $s$, we will show that its spin polarization parameter space is complex projective space $\mathbb{CP}^{2s}$. For example, the parameter spaces for a fermion $f$ and a massive gauge boson $A$ are $\mathbb{CP}^{1}$ and $\mathbb{CP}^{2}$, respectively. Moreover, we do not distinguish the fermion ($f$) and anti-fermion ($\bar{f}$) in the general discussions, similar to the supersymmetry theory. Also, we only consider the massive gauge bosons since they need to decay.

For a general quantum system with $N$ particles with spin $s_1$, $s_2$, ..., $s_N$, the quantum space is complex projective space  $\mathbb{CP}^{J-1}$ with $J=(2s_1+1) \times (2s_2+1) \times ... \times (2s_N+1)$, and the classical space is the cartesian product of the complex projective spaces $\mathbb{CP}^{2s_1} \times \mathbb{CP}^{2s_2} \times ... \times \mathbb{CP}^{2s_N}$. In mathematics, the classical space is the (generalized) Segre variety in the quantum space. For the $ff$, $AA$, $Af$, $fff$, and $ffA$ systems which can be produced at the colliders in high energy physics experiments, we shall calculate their discriminants $\Delta_i$, which are degree 2 homogeneous and holomorphic functions. We define the corresponding classical spaces as the discriminant locus $\Delta=0 $ for $ff$ system, and the intersections of the discriminant loci $\Delta_i=0$ for $AA$, $Af$, $fff$, and $ffA$ systems in the quantum space. Therefore, we define the quantum entanglement spaces as the quantum space with $\Delta\not=0 $ for $ff$ system, and the quantum spaces without the intersections of the discriminant loci $\Delta_i=0$ for $AA$, $Af$, $fff$, and $ffA$ systems. For two fermion $ff$ system, we prove that our discriminant criterion is equivalent to the original Peres-Horodecki criterion~\cite{Peres:1996dw, Horodecki:1997vt} and the CHSH criterion~\cite{Clauser:1969ny, Horodecki:1995nsk}. 
Thus, our discriminant criterion for classical
space is the same as the original Peres-Horodecki criterion, and is the same as 
the CHSH criterion for the Bell local parameter space. Also, 
our discriminant criterion for quantum entanglement space is the same as the original Peres-Horodecki criterion, and is the same as the CHSH
criterion for the Bell non-local parameter space. Therefore, we prove that our classical
space is the same as the Bell local parameter space, and our quantum entanglement
space is the same as the Bell non-local parameter space. In particular, our quantum
entanglement space is Bell non-local in high energy physics.
For high energy physics experiments such as colliders, we can reconstruct the discriminants from various measurements, and probe the quantum entanglement spaces at exact level.  We can perform such kind of studies in some two-fermion systems, but in general it might be very difficult. For classification, this kind of quantum entanglement search can be defined as the fundamental approach, or say kinematic approach. 
By the way, for real case, we obtain  $\Delta=-\frac{1}{2}C_{22}$. 

The other approach to probe the quantum entanglement parameter spaces is the specific approach, or say decay approach. The principle is the same. For any physics observable, we can calculate its quantum range and classical range for its expectation value, and the difference is the quantum entanglement range. In particular,
we develop a universal framework for identifying entanglement in high-energy systems by exploiting angular correlations in decay products. For $t\bar{t}$, azimuthal observables such as $D' = \frac{32}{\pi^2}\langle \cos(\phi_{e^+} - \phi_{e^-}) \rangle$ are shown to resolve entanglement where $D$ fails, with $D' \in [-1, -\frac{1}{2}) \cup (\frac{1}{2}, 1]$ serving as a robust criterion. Beyond the Standard Model (BSM), scenarios involving Higgs-like scalars ($h'$) or heavy $b'$ quarks enhance entanglement through distinct angular correlations—for instance, $h' \to W^{-}W^{+}$ decays ($m_{h'} < 254.1$ GeV) yield $\langle\cos\theta_{e^{-}e^{+}}\rangle > \frac{1}{4}$, while $b' \to W^{-}t$ decays ($m_{b'} < 319.5$ GeV) produce $D'_{W^{-}t} =\frac{2}{C_t} \langle\cos\theta_{e^{-}W^{+}}\rangle > 1$. For multi-body systems like $ttt$ and $t\bar{t}W^{-}$, triple-product correlations such as $\langle (\hat{e}_1 \times \hat{e}_2) \cdot \hat{e}_3 \rangle$ are introduced as novel probes of genuine tripartite entanglement.  

By bridging quantum information principles with collider phenomenology, we propose a brand new formalism for quantum entanglement, and its generic searches in high energy physics. In other words, our studies
establish the model-independent and solid methodologies to uncover entanglement-driven signatures. These results not only advance tests of quantum mechanics at high energies but also provide new strategies to search for exotic particles. Our framework paves the way for experimental validation at future $e^+e^-$ colliders and sets the stage for extending entanglement studies to complex systems, where quantum correlations may reveal deeper insights into the SM and beyond.  

The remainder of this paper is structured as follows. We propose a quantum entanglement theory for multi-particle system, and a fundamental approach to probe the quantum entanglement in Section~\ref{QET}.
Section~\ref{sec:15} develops a unified theoretical framework to detect the quantum entanglement in collider-produced multi-particle systems via a specific approach or decay approach. We derive the model-independent observables for multi-particle systems by employing the angular correlations and phase-space integration. Section~\ref{sec:2} revisits $t\bar{t}$ entanglement at $e^{+}e^{-}$ colliders, demonstrating the failure of the conventional observable $D = -3\langle\cos\theta_{e^{+}e^{-}}\rangle$ within the SM, and introduces the azimuthal observable $D^{\prime}$ as robust alternatives. Section~\ref{sec16} generalizes these methodologies to diverse systems, establishing entanglement criteria and highlighting sensitivity to BSM physics. Section~\ref{sec:7} concludes with implications for collider experiments, emphasizing applications in the SM tests, exotic particle searches, and future extensions of this framework.

\section{Quantum Entanglement Theory}\label{QET}

First, let us briefly review the mathematics background. The complex projective space $\mathbb{CP}^{n}$ is the $(n+1)$-dimensional complex space $\mathbb{C}^{n+1}\backslash \{0\}$ modulo the following equivalent classes
\begin{align}  
z \sim w ~~{\rm iff}~~\exists \lambda \in \mathbb{C}\backslash \{0\}, ~w=\lambda z~,~
\end{align} 
or explicit speaking
\begin{align}  
(z_0, z_1, z_2, \cdot \cdot \cdot, z_{n}) \sim (\lambda z_0, \lambda z_1, \lambda z_2, \cdot \cdot \cdot, \lambda z_{n}) ~~{\rm for~all}~~ \lambda \in \mathbb{C}\backslash \{0\}~.~
\end{align} 
In physics, we usually have $|z|=1$, and then have $|\lambda| =1$ as well. Obviously, the complex dimension of $\mathbb{CP}^{n}$ is $n$, or say the real dimension of $\mathbb{CP}^{n}$ is $2n$. Moreover, we have 
$\mathbb{CP}^{n} \simeq S^{2n+1}/ S^1$, where $S^n$ is $n$-dimensional sphere. In particular, we can prove that  
$\mathbb{CP}^{1}$ is diffeomorphic to $S^2$.

For a particle $P$ with spin $s$, the polarization state for its spin (or helicity) space is given by 
\begin{align}  
    |P\rangle = \sum^{s}_{i=-s} z_i |i\rangle ~,~
\end{align} 
where the normalization condition is 
\begin{align}  
    \sum^{s}_{i=-s} |z_i|^2 =1~.~ 
\end{align} 
Moreover, two polarization states are equivalent if their coefficients $z_i$ and $z_i'$ satisfy the following relation
\begin{align}  
(z_{-s}, z_{-s+1}, \cdot \cdot \cdot, z_{s}) = (\lambda z'_{-s}, \lambda z'_{-s+1}, \cdot \cdot \cdot, \lambda z'_{s}) ~,\quad \lambda \in \mathbb{C}~~{\rm and}~~|\lambda|=1~.~
\end{align}  
Therefore, we prove that the spin polarization space for particle $P$ with spin $s$ is $\mathbb{CP}^{2s}$.

For a general quantum system with $N$ particles with spin $s_1$, $s_2$, $\cdot \cdot \cdot$, and $s_N$, we obtain that the quantum space is complex projective space  $\mathbb{CP}^{J-1}$ with $J=\Pi_i (2s_i+1)=(2s_1+1) \times (2s_2+1) \times \cdot \cdot \cdot \times (2s_N+1)$, and the classical space is the cartesian product of the complex projective spaces $\mathbb{CP}^{2s_1} \times \mathbb{CP}^{2s_2} \times \cdot \cdot \cdot \times \mathbb{CP}^{2s_N}$. 
In mathematics, the classical space is the (generalized) Segre variety in the quantum space.
To define the quantum entanglement space, we propose the discriminants $\Delta_i$, which are degree 2 homogeneous and holomorphic functions. Thus, the Number of Independent Discriminants (NID) is 
\begin{align}  
{\rm NID}=J-1-\sum_i 2s_i= \Pi_i (2s_i+1)-1 
-\sum_i 2s_i~.~\label{nnid}
\end{align}  

We define the corresponding classical spaces as the discriminant locus $\Delta=0 $ for $ff$ system, and the intersections of the discriminant loci $\Delta_i=0$ for all the other systems. Therefore, we define the quantum entanglement spaces as the quantum space with $\Delta\not=0 $ for $ff$ system, and the quantum spaces without the intersections of the discriminant loci $\Delta_i=0$ for all the other systems.

For high energy physics experiments, we can reconstruct the discriminants from various measurements, and probe the quantum entanglement spaces at exact level.  We can perform such kind of studies in some two-fermion systems, but in general it might be very difficult. For classification, this kind of quantum entanglement search can be defined as the fundamental approach, or say kinematic approach. Moreover, to probe the quantum no-locality, we just consider the space-like separated measurements.

We shall study the discriminants in the $ff$, $AA$, $Af$, $fff$, and $ffA$ systems, as well as the general systems with two particles and three particles in the following subsections. To be general, we will not distinguish the fermion ($f$) and anti-fermion ($\bar{f}$), similar to the supersymmetry theory. Also, we only consider the massive gauge bosons since they need to decay.

\subsection{The Two-Fermion $f_1f_2$ System  }

The most general polarization state of a two-fermion system $f_1f_2$ can be written as 
\begin{align}  
    |f_1 f_2\rangle = \sum_{k,j=\pm {1\over 2}}\alpha_{k,j} |k\rangle_{f_1} \otimes |j\rangle_{f_2}
\end{align}  
with the normalization condition  
\begin{align}  
    \sum_{k,j=\pm {1\over 2}}|\alpha_{k,j}|^2 = 1~.  
\end{align}  

In the two-fermion system $f_1f_2$, the quantum space is $\mathbb{CP}^3$ with complex dimension 3, and the classical space is $\mathbb{CP}^1\otimes \mathbb{CP}^1$ with complex dimension 2. Thus, there is one discriminant with complex dimension 1, and we define it as
\begin{align} 
\Delta =  \alpha_{\frac{1}{2},\frac{1}{2}}\alpha_{-\frac{1}{2},-\frac{1}{2}}-\alpha_{\frac{1}{2},-\frac{1}{2}}\alpha_{-\frac{1}{2},\frac{1}{2}}~.~
\end{align}  
So the discriminant $\Delta$ is a degree 2 homogeneous and holomorphic function. Also, we can prove the following in-equalities
\begin{align} 
\left|\Delta\right| \leq & |\alpha_{\frac{1}{2},\frac{1}{2}}||\alpha_{-\frac{1}{2},-\frac{1}{2}}|+|\alpha_{\frac{1}{2},-\frac{1}{2}}||\alpha_{-\frac{1}{2},\frac{1}{2}}|
\nonumber\\  
   \leq & \frac{1}{2} \left( |\alpha_{\frac{1}{2},\frac{1}{2}}|^2+|\alpha_{-\frac{1}{2},-\frac{1}{2}}|^2+|\alpha_{\frac{1}{2},-\frac{1}{2}}|^2|\alpha_{-\frac{1}{2},\frac{1}{2}}|^2\right)=\frac{1}{2} ~.~
\end{align}
Thus, we obtain the range of $\left|\Delta\right|$
\begin{align} 
 0 \leq \left|\Delta\right| \leq \frac{1}{2}~.~
\end{align}
In addition, we define
\begin{align} 
&\alpha_{\frac{1}{2},\frac{1}{2}} \equiv |\alpha_{\frac{1}{2},\frac{1}{2}}| e^{i\eta_{++}}~,~ \alpha_{\frac{1}{2},-\frac{1}{2}} \equiv |\alpha_{\frac{1}{2},-\frac{1}{2}}| e^{i\eta_{+-}}~,~\nonumber\\ &
\alpha_{-\frac{1}{2},\frac{1}{2}} \equiv |\alpha_{-\frac{1}{2},\frac{1}{2}}| e^{i\eta_{-+}}~,~ \alpha_{-\frac{1}{2},-\frac{1}{2}} \equiv |\alpha_{-\frac{1}{2},-\frac{1}{2}}| e^{i\eta_{--}}~.~
\end{align}
We can rewrite the determinant $\Delta$ as two real functions $\Delta_M$ and $\Delta_P$ as below
\begin{align} 
\Delta_M = |\alpha_{\frac{1}{2},\frac{1}{2}}||\alpha_{-\frac{1}{2},-\frac{1}{2}}|-|\alpha_{\frac{1}{2},-\frac{1}{2}}||\alpha_{-\frac{1}{2},\frac{1}{2}}|~,~
\end{align}
\begin{align} 
\Delta_P = \eta_{++} + \eta_{--} - \eta_{+-} - \eta_{-+} ~.~
\end{align}
The classical space is the discriminant locus $\Delta=0$ in quantum space, or say the quantum space with both
$\Delta_M=0$ and $\Delta_P=0$. And thus the quantum entanglement space is the quantum space with $\Delta\not=0$, or $\Delta_M\not=0$, or $\Delta_P\not=0$.
In particular, even if $\Delta_M=0$, we might still probe the quantum entanglement space with $\Delta_P\not=0$. Moreover, we can reconstruct $\Delta$ from the collider experiments, for example, the $\Lambda \overline{\Lambda}$ pair productions at the BES experiment \cite{Pei:2025yvr}. Thus, we can probe the quantum entanglement space via the fundamental approach. If the measurements are space-like separated, we can probe the quantum non-locality as well.

First, we study the relation between our discriminant and the Peres-Horodecki criterion~\cite{Peres:1996dw, Horodecki:1997vt}. For simplicity, we follow the discussions in Appendix A in Ref.~\cite{Afik:2020onf}.
In the basis $\left(|\frac{1}{2}\rangle_{f_1} \otimes |\frac{1}{2}\rangle_{f_2}, |\frac{1}{2}\rangle_{f_1} \otimes |-\frac{1}{2}\rangle_{f_2}, |-\frac{1}{2}\rangle_{f_1} \otimes |\frac{1}{2}\rangle_{f_2}, |-\frac{1}{2}\rangle_{f_1} \otimes |-\frac{1}{2}\rangle_{f_2} \right)$, we obtain the spin density matrix
\begin{align}
    \rho=|f_1 f_2\rangle \langle f_1 f_2|=
    \begin{pmatrix}
    |\alpha_{\frac{1}{2},\frac{1}{2}}|^2 
    & \alpha_{\frac{1}{2},\frac{1}{2}}\alpha^*_{\frac{1}{2},-\frac{1}{2}} 
    & \alpha_{\frac{1}{2},\frac{1}{2}}\alpha^*_{-\frac{1}{2},\frac{1}{2}} 
    & \alpha_{\frac{1}{2},\frac{1}{2}}\alpha^*_{-\frac{1}{2},-\frac{1}{2}}  \\ 
     \alpha_{\frac{1}{2},-\frac{1}{2}}\alpha^*_{\frac{1}{2},\frac{1}{2}} 
    & |\alpha_{\frac{1}{2},-\frac{1}{2}}|^2 
    & \alpha_{\frac{1}{2},-\frac{1}{2}}\alpha^*_{-\frac{1}{2},\frac{1}{2}} 
    & \alpha_{\frac{1}{2},-\frac{1}{2}}\alpha^*_{-\frac{1}{2},-\frac{1}{2}}  \\
    \alpha_{-\frac{1}{2},\frac{1}{2}}\alpha^*_{\frac{1}{2},\frac{1}{2}} 
    & \alpha_{-\frac{1}{2},\frac{1}{2}}\alpha^*_{\frac{1}{2},-\frac{1}{2}} 
    & |\alpha_{-\frac{1}{2},\frac{1}{2}}|^2
    & \alpha_{-\frac{1}{2},\frac{1}{2}}\alpha^*_{-\frac{1}{2},-\frac{1}{2}}  \\
    \alpha_{-\frac{1}{2},-\frac{1}{2}}\alpha^*_{\frac{1}{2},\frac{1}{2}} 
    & \alpha_{-\frac{1}{2},-\frac{1}{2}}\alpha^*_{\frac{1}{2},-\frac{1}{2}} 
    & \alpha_{-\frac{1}{2},-\frac{1}{2}}\alpha^*_{-\frac{1}{2},\frac{1}{2}}
    & |\alpha_{-\frac{1}{2},-\frac{1}{2}}|^2
    \end{pmatrix}.\label{rhoff}
\end{align}
Taking partial transpose of $\rho$, {\it i.e.},
the transposes of the four $2\times 2$ sub-matrices of $\rho$, we obtain
\begin{align}
    \rho^{\rm{T}_2}=
    \begin{pmatrix}
    |\alpha_{\frac{1}{2},\frac{1}{2}}|^2 
    & \alpha_{\frac{1}{2},-\frac{1}{2}}\alpha^*_{\frac{1}{2},\frac{1}{2}} 
    & \alpha_{\frac{1}{2},\frac{1}{2}}\alpha^*_{-\frac{1}{2},\frac{1}{2}} 
    & \alpha_{\frac{1}{2},-\frac{1}{2}}\alpha^*_{-\frac{1}{2},\frac{1}{2}} 
    \\ 
     \alpha_{\frac{1}{2},\frac{1}{2}}\alpha^*_{\frac{1}{2},-\frac{1}{2}} 
    & |\alpha_{\frac{1}{2},-\frac{1}{2}}|^2 
    & \alpha_{\frac{1}{2},\frac{1}{2}}\alpha^*_{-\frac{1}{2},-\frac{1}{2}} 
    & \alpha_{\frac{1}{2},-\frac{1}{2}}\alpha^*_{-\frac{1}{2},-\frac{1}{2}}  \\
    \alpha_{-\frac{1}{2},\frac{1}{2}}\alpha^*_{\frac{1}{2},\frac{1}{2}} 
    & \alpha_{-\frac{1}{2},-\frac{1}{2}}\alpha^*_{\frac{1}{2},\frac{1}{2}} 
    & |\alpha_{-\frac{1}{2},\frac{1}{2}}|^2
    & \alpha_{-\frac{1}{2},-\frac{1}{2}}\alpha^*_{-\frac{1}{2},\frac{1}{2}} 
    \\
    \alpha_{-\frac{1}{2},\frac{1}{2}}\alpha^*_{\frac{1}{2},-\frac{1}{2}} 
    & \alpha_{-\frac{1}{2},-\frac{1}{2}}\alpha^*_{\frac{1}{2},-\frac{1}{2}} 
    & \alpha_{-\frac{1}{2},\frac{1}{2}}\alpha^*_{-\frac{1}{2},-\frac{1}{2}}  
    & |\alpha_{-\frac{1}{2},-\frac{1}{2}}|^2
    \end{pmatrix}.\label{rhoff-T2}
\end{align}
The original  Peres-Horodecki criterion provides a sufficient and necessary condition for classical space: $\rho^{\rm{T}_2}$ is positive semi-definite. The four eigenvalues of $\rho^{\rm{T}_2}$ are
\begin{align} 
-|\Delta|~, ~~|\Delta|~, ~~ \frac{1}{2}\left(1-\sqrt{1-4|\Delta|^2}\right)~, ~~ \frac{1}{2}\left(1+\sqrt{1-4|\Delta|^2}\right)~.~
\end{align}  
Thus, the original  Peres-Horodecki criterion for classical space is $\Delta=0$. Therefore, we prove that our criterion for classical space is equivalent to the
original  Peres-Horodecki criterion.

In addition, we have  
\begin{align}  
  C_{22} &= \rho_{23}+ \rho_{32}- \rho_{14}-\rho_{41} 
  \nonumber \\ &
  = \alpha_{\frac{1}{2},-\frac{1}{2}}\alpha^*_{-\frac{1}{2},\frac{1}{2}} + \alpha_{-\frac{1}{2},\frac{1}{2}}\alpha^*_{\frac{1}{2},-\frac{1}{2}} 
  - \alpha_{\frac{1}{2},\frac{1}{2}}\alpha^*_{-\frac{1}{2},-\frac{1}{2}} - \alpha_{-\frac{1}{2},-\frac{1}{2}}\alpha^*_{\frac{1}{2},\frac{1}{2}}  ~.~
\end{align}  
If $\alpha_{\frac{1}{2},\frac{1}{2}}$, $\alpha_{\frac{1}{2},-\frac{1}{2}}$, $\alpha_{-\frac{1}{2},\frac{1}{2}}$, and $\alpha_{-\frac{1}{2},-\frac{1}{2}}$ are all real for real case, we obtain
\begin{align}  
\Delta = -\frac{1}{2} C_{22}~.~
\end{align}

Second, Bell inequality is an equation which distinguishes the Bell local quantum states (or general speaking Bell local states)
and the Bell non-local quantum states. For a bipartite qubit system, it is the CHSH inequality~\cite{Clauser:1969ny}. Thus, we study the relation between our discriminant criterion and the CHSH inequality.
For simplicity, we consider the equivalent definition of the CHSH inequality in Ref.~\cite{Horodecki:1995nsk}.

With $\rho$ defined in Eq.~(\ref{rhoff}) and Eq.~(\ref{rho1}), we can rewrite $B^{\pm}_i$ and $C_{ij}$ in terms of $\alpha_{k,j}$ and $\alpha^*_{k, j}$ as follows
\begin{align}
    & B^+_1=\alpha_{\frac{1}{2},\frac{1}{2}} \alpha^*_{-\frac{1}{2},\frac{1}{2}}+\alpha_{\frac{1}{2},-\frac{1}{2}} \alpha^*_{-\frac{1}{2},-\frac{1}{2}}+\alpha_{-\frac{1}{2},\frac{1}{2}} \alpha^*_{\frac{1}{2},\frac{1}{2}}+\alpha_{-\frac{1}{2},-\frac{1}{2}} \alpha^*_{\frac{1}{2},-\frac{1}{2}}~,\label{bb1}\\
    & B^+_2=i (\alpha_{\frac{1}{2},\frac{1}{2}} \alpha^*_{-\frac{1}{2},\frac{1}{2}}+\alpha_{\frac{1}{2},-\frac{1}{2}} \alpha^*_{-\frac{1}{2},-\frac{1}{2}}-\alpha_{-\frac{1}{2},\frac{1}{2}} \alpha^*_{\frac{1}{2},\frac{1}{2}}-\alpha_{-\frac{1}{2},-\frac{1}{2}} \alpha^*_{\frac{1}{2},-\frac{1}{2}})~,\\
    & B^+_3=2 \alpha_{\frac{1}{2},\frac{1}{2}} \alpha^*_{\frac{1}{2},\frac{1}{2}}+2 \alpha_{\frac{1}{2},-\frac{1}{2}} \alpha^*_{\frac{1}{2},-\frac{1}{2}}-1~,\\
    & B^-_1=\alpha_{\frac{1}{2},\frac{1}{2}} \alpha^*_{\frac{1}{2},-\frac{1}{2}}+\alpha_{\frac{1}{2},-\frac{1}{2}} \alpha^*_{\frac{1}{2},\frac{1}{2}}+\alpha_{-\frac{1}{2},\frac{1}{2}} \alpha^*_{-\frac{1}{2},-\frac{1}{2}}+\alpha_{-\frac{1}{2},-\frac{1}{2}} \alpha^*_{-\frac{1}{2},\frac{1}{2}}~,\\
    & B^-_2=i (\alpha_{\frac{1}{2},\frac{1}{2}} \alpha^*_{\frac{1}{2},-\frac{1}{2}}-\alpha_{\frac{1}{2},-\frac{1}{2}} \alpha^*_{\frac{1}{2},\frac{1}{2}}+\alpha_{-\frac{1}{2},\frac{1}{2}} \alpha^*_{-\frac{1}{2},-\frac{1}{2}}-\alpha_{-\frac{1}{2},-\frac{1}{2}} \alpha^*_{-\frac{1}{2},\frac{1}{2}})~,\\
    & B^-_3=2 \alpha_{\frac{1}{2},\frac{1}{2}} \alpha^*_{\frac{1}{2},\frac{1}{2}}+2 \alpha_{-\frac{1}{2},\frac{1}{2}} \alpha^*_{-\frac{1}{2},\frac{1}{2}}-1~,\\
    & C_{11}= \alpha_{\frac{1}{2},\frac{1}{2}} \alpha^*_{-\frac{1}{2},-\frac{1}{2}}+\alpha_{\frac{1}{2},-\frac{1}{2}} \alpha^*_{-\frac{1}{2},\frac{1}{2}}+\alpha_{-\frac{1}{2},\frac{1}{2}} \alpha^*_{\frac{1}{2},-\frac{1}{2}}+\alpha_{-\frac{1}{2},-\frac{1}{2}} \alpha^*_{\frac{1}{2},\frac{1}{2}}~,\\
    & C_{12}= i (\alpha_{\frac{1}{2},\frac{1}{2}} \alpha^*_{-\frac{1}{2},-\frac{1}{2}}-\alpha_{\frac{1}{2},-\frac{1}{2}} \alpha^*_{-\frac{1}{2},\frac{1}{2}}+\alpha_{-\frac{1}{2},\frac{1}{2}} \alpha^*_{\frac{1}{2},-\frac{1}{2}}-\alpha_{-\frac{1}{2},-\frac{1}{2}} \alpha^*_{\frac{1}{2},\frac{1}{2}})~,\\
    & C_{13}=\alpha_{\frac{1}{2},\frac{1}{2}} \alpha^*_{-\frac{1}{2},\frac{1}{2}}-\alpha_{\frac{1}{2},-\frac{1}{2}} \alpha^*_{-\frac{1}{2},-\frac{1}{2}}+\alpha_{-\frac{1}{2},\frac{1}{2}} \alpha^*_{\frac{1}{2},\frac{1}{2}}-\alpha_{-\frac{1}{2},-\frac{1}{2}} \alpha^*_{\frac{1}{2},-\frac{1}{2}}~,\\
    & C_{21}= i (\alpha_{\frac{1}{2},\frac{1}{2}} \alpha^*_{-\frac{1}{2},-\frac{1}{2}}+\alpha_{\frac{1}{2},-\frac{1}{2}} \alpha^*_{-\frac{1}{2},\frac{1}{2}}-\alpha_{-\frac{1}{2},\frac{1}{2}} \alpha^*_{\frac{1}{2},-\frac{1}{2}}-\alpha_{-\frac{1}{2},-\frac{1}{2}} \alpha^*_{\frac{1}{2},\frac{1}{2}})~,\\
    & C_{22}= -\alpha_{\frac{1}{2},\frac{1}{2}} \alpha^*_{-\frac{1}{2},-\frac{1}{2}}+\alpha_{\frac{1}{2},-\frac{1}{2}} \alpha^*_{-\frac{1}{2},\frac{1}{2}}+\alpha_{-\frac{1}{2},\frac{1}{2}} \alpha^*_{\frac{1}{2},-\frac{1}{2}}-\alpha_{-\frac{1}{2},-\frac{1}{2}} \alpha^*_{\frac{1}{2},\frac{1}{2}}~,\\
    & C_{23}=i (\alpha_{\frac{1}{2},\frac{1}{2}} \alpha^*_{-\frac{1}{2},\frac{1}{2}}-\alpha_{\frac{1}{2},-\frac{1}{2}} \alpha^*_{-\frac{1}{2},-\frac{1}{2}}-\alpha_{-\frac{1}{2},\frac{1}{2}} \alpha^*_{\frac{1}{2},\frac{1}{2}}+\alpha_{-\frac{1}{2},-\frac{1}{2}} \alpha^*_{\frac{1}{2},-\frac{1}{2}})~,\\
    & C_{31}= \alpha_{\frac{1}{2},\frac{1}{2}} \alpha^*_{\frac{1}{2},-\frac{1}{2}}+\alpha_{\frac{1}{2},-\frac{1}{2}} \alpha^*_{\frac{1}{2},\frac{1}{2}}-\alpha_{-\frac{1}{2},\frac{1}{2}} \alpha^*_{-\frac{1}{2},-\frac{1}{2}}-\alpha_{-\frac{1}{2},-\frac{1}{2}} \alpha^*_{-\frac{1}{2},\frac{1}{2}}~,\\
    & C_{32}= i (\alpha_{\frac{1}{2},\frac{1}{2}} \alpha^*_{\frac{1}{2},-\frac{1}{2}}-\alpha_{\frac{1}{2},-\frac{1}{2}} \alpha^*_{\frac{1}{2},\frac{1}{2}}-\alpha_{-\frac{1}{2},\frac{1}{2}} \alpha^*_{-\frac{1}{2},-\frac{1}{2}}+\alpha_{-\frac{1}{2},-\frac{1}{2}} \alpha^*_{-\frac{1}{2},\frac{1}{2}})~,\\
    & C_{33}=1-2 \alpha_{\frac{1}{2},-\frac{1}{2}} \alpha^*_{\frac{1}{2},-\frac{1}{2}}-2 \alpha_{-\frac{1}{2},\frac{1}{2}} \alpha^*_{-\frac{1}{2},\frac{1}{2}}~.\label{bbn}
\end{align}
Thus, if $\alpha_{k,j}$ are all real for real case, we obtain 
\begin{align}
C_{12}=C_{21}=C_{23}=C_{32}=0~.~
\end{align}

Defining the matrix $C$ as the matrix with elements $C_{ij}$, we obtain 
the eigenvalues of $C^\text{T}C$ 
\begin{align}
    \lambda_1=1~,\quad \lambda_2=\lambda_3=4\left|\Delta\right|^2~.
\end{align}
Note that $0\le\left|\Delta\right|\le \frac{1}{2}$, the ranges of $\lambda_{2,3}$ are given by 
\begin{align}
    0\le\lambda_{2,3}\le 1~.
\end{align}
Thus, we obtain the Bell variable  for the CHSH inequality~\cite{Clauser:1969ny, Horodecki:1995nsk}
\begin{align}
{\cal{B}} \equiv 2\sqrt{1+4|\Delta|^2}~.~
\end{align}
Because $0\leq |\Delta| \leq \frac{1}{2}$, we prove 
\begin{align}
2 \leq {\cal{B}} \leq 2{\sqrt 2}~.~
\end{align}
The CHSH inequality is $ {\cal{B}} \leq 2$, and thus for the Bell local parameter space the CHSH inequality becomes the  CHSH criterion $ {\cal{B}} = 2$, {\it i.e.,} $\Delta = 0$.
The CHSH inequality is violated if and only if $ {\cal{B}} > 2$, {\it i.e.,} $\Delta \not= 0$.
Thus, our discriminant criterion for classical space is the same as the CHSH criterion for
the Bell local parameter space, and our discriminant criterion 
for quantum entanglement space is 
the same as the CHSH criterion for the Bell non-local parameter space. Therefore,
we prove that our classical space is the same as the Bell local parameter space,
and our  quantum entanglement space is the same as the Bell non-local parameter space. 
In particular, our quantum entanglement space is Bell non-local in high energy physics.
And then we do not need to mention the Bell non-locality for the quantum entanglement searches 
in the high energy physics experiments.

To distinguish the classical space and Bell local parameter space, or distinguish 
the quantum entanglement space and Bell non-local parameter space, we consider 
the  Werner states~\cite{Werner:1989zz}.
Because our discriminant criterion is the same as the CHSH criterion, we need to prove that 
the Werner state, which satisfies the Peres-Horodecki criterion for quantum entanglement  space
and the CHSH criterion for Bell local parameter space, 
does not exist in our quantum entanglement space.

The Werner states with a free parameter $w$ for the spin density matrix are
\begin{align}
    \rho_{\rm W}=
    \begin{pmatrix}
    \frac{1-w}{4} & 0  & 0 & 0  \\ 
     0 &  \frac{1+w}{4} &  -\frac{w}{2} & 0  \\
    0  & -\frac{w}{2}  & \frac{1+w}{4} & 0 \\
    0 & 0 & 0 & \frac{1-w}{4}
    \end{pmatrix}~,~ w \in \left[-\frac{1}{3}, ~1 \right]~.~
    \label{rho-werner}
\end{align}
Using the Peres-Horodecki criterion~\cite{Peres:1996dw, Horodecki:1997vt}, we can prove that 
the Werner states are classical or say
separable for $w\leq \frac{1}{3}$, and quantum entangled for $w > \frac{1}{3}$. In addition, 
the Bell variable for the
CHSH inequality for a Werner state is $2\sqrt{2} \left|w\right|$, {\it i.e.}, ${\cal{B}} =2\sqrt{2} \left|w\right|$~\cite{Clauser:1969ny, Horodecki:1995nsk}. And thus 
we can realize Bell non-locality for $w> \frac{1}{\sqrt 2}$. Therefore, for the Werner states, the quantum entanglement range is
 $w \in \left(\frac{1}{3}, ~1 \right]$, and the Bell non-locality range is 
 $w \in \left(\frac{1}{\sqrt 2}, ~1 \right]$. However, comparing with the spin density matrix 
 in Eq.~(\ref{rhoff}), we can easily show that the Werner states with 
 $w \in \left(\frac{1}{3}, ~ \frac{1}{\sqrt 2}\right]$ cannot be realized. In fact,
 we can only achieve the Werner state with $w=1$.
Therefore, we prove that 
the Werner state, which satisfies the Peres-Horodecki criterion for quantum entanglement  space and the CHSH criterion for Bell local parameter space, 
does not exist in our quantum entanglement space.

By the way, if $\alpha_{k,j}$ are all real for real case, we obtain 
\begin{align}
{\cal{B}} \equiv 2\sqrt{1+C_{22}^2} ~.~
\end{align}
This Bell variable has been realized in the literature, for example, Refs.~\cite{Wu:2024asu, Han:2025ewp}.

\subsection{The Two-Gauge Boson $AA$ System }

We emphasize that we only consider the massive gauge bosons since they need to decay.
The most general polarization state of a physical system with two gauge bosons $A_1A_2$ can be written as 
\begin{align}  
    |A_1 A_2\rangle = \sum_{j,k=1, 0, -1} \alpha_{j,k} |j\rangle_{A_1} \otimes |k\rangle_{A_2}
\end{align}  
with the normalization condition  
\begin{align}  
    \sum_{j,k=1, 0, -1}|\alpha_{j,k}|^2 = 1~.  
\end{align}  

In the two-gauge boson system $A_1A_2$, the quantum space is $\mathbb{CP}^8$ with complex dimension 8, and the classical space is $\mathbb{CP}^2\otimes \mathbb{CP}^2$ with complex dimension 4. Thus, there are four independent discriminants with complex dimension 1. We define the general discriminants as
\begin{equation}
    \begin{split}
        & \Delta_1 =  \alpha_{1,1} \alpha_{0,0}-\alpha_{1,0} \alpha_{0,1}~,~\\ 
        & \Delta_2 =  \alpha_{1,1} \alpha_{0,-1}-\alpha_{1,-1} \alpha_{0,1}~,~\\ 
        & \Delta_3 =  \alpha_{1,1} \alpha_{-1,0}-\alpha_{1,0} \alpha_{-1,1}~,~\\ 
        & \Delta_4 =  \alpha_{1,1} \alpha_{-1,-1}-\alpha_{1,-1} \alpha_{-1,1}~,~\\
        & \Delta_5 =  \alpha_{-1,-1} \alpha_{0,0}-\alpha_{-1,0} \alpha_{0,-1}~,~ \\
        & \Delta_6 =  \alpha_{-1,-1} \alpha_{0,1}-\alpha_{-1,1} \alpha_{0,-1}~,~ \\
        & \Delta_7 =  \alpha_{-1,-1} \alpha_{1,0}-\alpha_{-1,0} \alpha_{1,-1}~,~ \\
        & \Delta_8 =  \alpha_{0,0} \alpha_{1,-1}-\alpha_{0,-1} \alpha_{1,0}~,~ \\
        & \Delta_9 =  \alpha_{0,0} \alpha_{-1,1}-\alpha_{0,1} \alpha_{-1,0}~.~
    \end{split}
\end{equation}
Thus, all the discriminants $\Delta_i$ are degree 2 homogeneous and holomorphic functions. 
Also, we can prove that the ranges of $\left|\Delta_i\right|$ are
\begin{align} 
 0 \leq \left|\Delta_i\right| \leq \frac{1}{2}~.~
\end{align}
In addition, with $\Delta_1$ and $\Delta_2$, we can obtain $\Delta_8$.
With $\Delta_1$ and $\Delta_3$, we can obtain $\Delta_9$.
With $\Delta_2$ and $\Delta_4$, we can obtain $\Delta_6$.
With $\Delta_3$ and $\Delta_4$, we can obtain $\Delta_7$.
And with $\Delta_7$ and $\Delta_8$, we can obtain $\Delta_5$.
Therefore, the four independent discriminants can be chosen as 
$\{\Delta_1, \Delta_2, \Delta_3, \Delta_4 \}$.
The classical space is the intersection of the discriminant loci $\Delta_i=0$ in quantum space, and the quantum entanglement space is the quantum space by removing
the intersection of the discriminant loci $\Delta_i=0$, {\it i.e.}, the quantum space minus the classical space.

\subsection{The Gauge Boson-Fermion $Af$ System }

The most general polarization state of a physics system with one gauge boson $A$ and one fermion $f$ can be written as 
\begin{align}  
    |A f\rangle = \sum_{j=1, 0, -1} \sum_{k=\pm\frac{1}{2}} \alpha_{j,k} |j\rangle_{A} \otimes |k\rangle_{f}
\end{align}  
with the normalization condition  
\begin{align}  
\sum_{j=1, 0, -1} \sum_{k=\pm\frac{1}{2}} |\alpha_{j,k}|^2 = 1~.  
\end{align}  

In the $Af$ system, the quantum space is $\mathbb{CP}^5$ with complex dimension 5, and the classical space is $\mathbb{CP}^2\otimes \mathbb{CP}^1$ with complex dimension 3. Thus, there are two independent discriminants with complex dimension 1. We define the general discriminants as
\begin{align} 
& \Delta_1 =  \alpha_{1,\frac{1}{2}} \alpha_{0,-\frac{1}{2}}-
\alpha_{1,-\frac{1}{2}} \alpha_{0,\frac{1}{2}}~,~
\nonumber \\ &
\Delta_2 =  \alpha_{1,\frac{1}{2}} \alpha_{-1,-\frac{1}{2}}-
\alpha_{1,-\frac{1}{2}} \alpha_{-1,\frac{1}{2}}~,~
\nonumber \\ &
\Delta_3 =  \alpha_{0,\frac{1}{2}} \alpha_{-1,-\frac{1}{2}}-
\alpha_{0,-\frac{1}{2}} \alpha_{-1,\frac{1}{2}}~.~
\end{align}  
Thus, all the discriminants $\Delta_i$ are degree 2 homogeneous and holomorphic functions. 
Also, we can prove that the ranges of $\left|\Delta_i\right|$ are
\begin{align} 
 0 \leq \left|\Delta_i\right| \leq \frac{1}{2}~.~
\end{align}
In addition, with $\Delta_1$ and $\Delta_2$, we can obtain $\Delta_3$.
Thus, the two independent discriminants can be chosen as 
$\{\Delta_1, \Delta_2\}$. The classical space is 
the intersection of the discriminant loci $\Delta_i=0$ in quantum space, and the quantum entanglement space is the quantum space without
the intersection of the discriminant loci $\Delta_i=0$, {\it i.e.}, the quantum space minus the classical space.

\subsection{The Three Fermion $fff$ System}

The most general polarization state of a physics system with three fermions $f_1 f_2 f_3$ can be written as 
\begin{align}  
    | f_1 f_2 f_3 \rangle = \sum_{j,k, l=\pm\frac{1}{2}} \alpha_{j,k, l} |j\rangle_{f_1} \otimes |k\rangle_{f_2} \otimes |l\rangle_{f_2}
\end{align}  
with the normalization condition  
\begin{align}  
    \sum_{j,k, l=\pm\frac{1}{2}}|\alpha_{j,k,l}|^2 = 1~.  
\end{align}  

In the physics system with three fermions $f_1 f_2 f_3$, the quantum space is $\mathbb{CP}^7$ with complex dimension 7, and the classical space is $\mathbb{CP}^1\otimes \mathbb{CP}^1 \otimes \mathbb{CP}^1$ with complex dimension 3. Thus, there are four independent discriminants with complex dimension 1. The strategy to construct the discriminants of $N+1$ particles is that we fix the spin (helicity) of one particle and construct the corresponding discriminants of $N$ particles, thus, we have $N+1$ kinds. Next, we consider the new discriminants where all the particles have different spins (helicities).
We define the general discriminants as
\begin{align} 
& \Delta_1 =  \alpha_{\frac{1}{2}, \frac{1}{2},\frac{1}{2}}\alpha_{\frac{1}{2}, -\frac{1}{2},-\frac{1}{2}} - \alpha_{\frac{1}{2}, \frac{1}{2},-\frac{1}{2}}\alpha_{\frac{1}{2}, -\frac{1}{2}, \frac{1}{2}}~,~
\nonumber \\ &
\Delta'_1 =  \alpha_{-\frac{1}{2}, \frac{1}{2},\frac{1}{2}}\alpha_{-\frac{1}{2}, -\frac{1}{2},-\frac{1}{2}}-\alpha_{-\frac{1}{2}, \frac{1}{2},-\frac{1}{2}}\alpha_{-\frac{1}{2}, -\frac{1}{2},\frac{1}{2}}~,~
\nonumber \\ &
\Delta_2 =  \alpha_{\frac{1}{2}, \frac{1}{2},\frac{1}{2}}\alpha_{-\frac{1}{2}, \frac{1}{2},-\frac{1}{2}} - \alpha_{\frac{1}{2}, \frac{1}{2},-\frac{1}{2}}\alpha_{-\frac{1}{2}, \frac{1}{2}, \frac{1}{2}}~,~
\nonumber \\ &
\Delta'_2 =  \alpha_{\frac{1}{2}, -\frac{1}{2},\frac{1}{2}}\alpha_{-\frac{1}{2}, -\frac{1}{2},-\frac{1}{2}} - \alpha_{\frac{1}{2}, -\frac{1}{2},-\frac{1}{2}}\alpha_{-\frac{1}{2}, -\frac{1}{2}, \frac{1}{2}}~,~ 
\nonumber \\ &
\Delta_3 =  \alpha_{\frac{1}{2}, \frac{1}{2},\frac{1}{2}}\alpha_{-\frac{1}{2}, -\frac{1}{2}, \frac{1}{2}} - \alpha_{\frac{1}{2}, -\frac{1}{2}, \frac{1}{2}}\alpha_{-\frac{1}{2}, \frac{1}{2}, \frac{1}{2}}~,~
\nonumber \\ &
\Delta'_3 =  \alpha_{\frac{1}{2}, \frac{1}{2},-\frac{1}{2}}\alpha_{-\frac{1}{2}, -\frac{1}{2}, -\frac{1}{2}} - \alpha_{\frac{1}{2}, -\frac{1}{2}, -\frac{1}{2}}\alpha_{-\frac{1}{2}, \frac{1}{2}, -\frac{1}{2}}~,~
\nonumber \\ &
\Delta_4 =  \alpha_{\frac{1}{2}, \frac{1}{2},\frac{1}{2}}\alpha_{-\frac{1}{2}, -\frac{1}{2},-\frac{1}{2}} - \alpha_{\frac{1}{2}, \frac{1}{2},-\frac{1}{2}}\alpha_{-\frac{1}{2}, -\frac{1}{2}, \frac{1}{2}}~,~
\nonumber \\ &
\Delta'_4 =  \alpha_{\frac{1}{2}, \frac{1}{2},\frac{1}{2}}\alpha_{-\frac{1}{2}, -\frac{1}{2},-\frac{1}{2}} - \alpha_{\frac{1}{2}, -\frac{1}{2},\frac{1}{2}}\alpha_{-\frac{1}{2}, \frac{1}{2}, -\frac{1}{2}}~,~
\nonumber \\ &
\Delta^{\prime \prime}_4 =  \alpha_{\frac{1}{2}, \frac{1}{2},\frac{1}{2}}\alpha_{-\frac{1}{2}, -\frac{1}{2},-\frac{1}{2}} - \alpha_{-\frac{1}{2}, \frac{1}{2},\frac{1}{2}}\alpha_{\frac{1}{2}, -\frac{1}{2}, -\frac{1}{2}}~.~
\end{align}  
Thus, all the discriminants $\Delta_i$ are degree 2 homogeneous and holomorphic functions. 
Also, we can prove that the ranges of $\left|\Delta_i\right|$ are
\begin{align} 
 0 \leq \left|\Delta_i\right| \leq \frac{1}{2}~.~
\end{align}
Moreover, with $\Delta_1$ and $\Delta_4$, we can obtain $\Delta'_2$.
With $\Delta_2$ and $\Delta_4$, we can obtain $\Delta'_1$.
With $\Delta_3$ and $\Delta'_1$, we can obtain $\Delta'_4$.
With $\Delta_1$ and $\Delta'_4$, we can obtain $\Delta'_3$.
And with $\Delta_2$ and $\Delta'_3$, we can obtain $\Delta^{\prime \prime}_4$.
Therefore, the four independent discriminants can be chosen as 
$\{\Delta_1, \Delta_2, \Delta_3, \Delta_4 \}$.
The classical space is the intersection of the discriminant loci $\Delta_i=0$ in quantum space, and the quantum entanglement space is the quantum space by removing
the intersection of the discriminant loci $\Delta_i=0$, {\it i.e.}, the quantum space minus the classical space.

\subsection{The System with Two Fermions and One Gauge Boson $ffA$ }

The most general polarization state of a physics system with two fermions and one gauge boson $f_1 f_2 A$ can be written as 
\begin{align}  
    | f_1 f_2 A \rangle = \sum_{j,k=\pm\frac{1}{2}} \sum_{l=1, 0, -1}\alpha_{j,k, l} |j\rangle_{f_1} \otimes |k\rangle_{f_2} \otimes |l\rangle_{A}
\end{align}  
with the normalization condition  
\begin{align}  
\sum_{j,k=\pm\frac{1}{2}} \sum_{l=1, 0, -1} |\alpha_{j,k, l}|^2=1~.~\,  
\end{align}  

In the physics system with two fermions and one gauge boson $f_1 f_2 A$, the quantum space is $\mathbb{CP}^{11}$ with complex dimension 11, and the classical space is $\mathbb{CP}^1\otimes \mathbb{CP}^1 \otimes \mathbb{CP}^2$ with complex dimension 4. Thus, there are seven independent discriminants with complex dimension 1. The strategy to construct the discriminants is similar to the three fermion system.
We define the general discriminants as
\begin{align} 
& \Delta_1 =  \alpha_{\frac{1}{2}, \frac{1}{2},1}\alpha_{\frac{1}{2}, -\frac{1}{2},0} - \alpha_{\frac{1}{2}, \frac{1}{2},0} \alpha_{\frac{1}{2}, -\frac{1}{2}, 1}~,~
\nonumber \\ &
\Delta'_1 =  \alpha_{-\frac{1}{2}, \frac{1}{2},1}\alpha_{-\frac{1}{2}, -\frac{1}{2},0} - \alpha_{-\frac{1}{2}, \frac{1}{2},0} \alpha_{-\frac{1}{2}, -\frac{1}{2}, 1}~,~
\nonumber \\ &
\Delta_2 =  \alpha_{\frac{1}{2}, \frac{1}{2},1}\alpha_{\frac{1}{2}, -\frac{1}{2},-1} - \alpha_{\frac{1}{2}, \frac{1}{2},-1} \alpha_{\frac{1}{2}, -\frac{1}{2}, 1}~,~
\nonumber \\ &
\Delta'_2 =  \alpha_{-\frac{1}{2}, \frac{1}{2},1}\alpha_{-\frac{1}{2}, -\frac{1}{2},-1} - \alpha_{-\frac{1}{2}, \frac{1}{2},-1} \alpha_{-\frac{1}{2}, -\frac{1}{2}, 1}~,~
\nonumber \\ &
\Delta_3 =  \alpha_{\frac{1}{2}, \frac{1}{2},0}\alpha_{\frac{1}{2}, -\frac{1}{2},-1} - \alpha_{\frac{1}{2}, \frac{1}{2},-1} \alpha_{\frac{1}{2}, -\frac{1}{2}, 0}~,~
\nonumber \\ &
\Delta'_3 =  \alpha_{-\frac{1}{2}, \frac{1}{2},0}\alpha_{-\frac{1}{2}, -\frac{1}{2},-1} - \alpha_{-\frac{1}{2}, \frac{1}{2},-1} \alpha_{-\frac{1}{2}, -\frac{1}{2}, 0}~,~
\nonumber \\ &
\Delta_4 =  \alpha_{\frac{1}{2}, \frac{1}{2},1}\alpha_{-\frac{1}{2}, \frac{1}{2},0} - \alpha_{\frac{1}{2}, \frac{1}{2},0} \alpha_{-\frac{1}{2}, \frac{1}{2}, 1}~,~
\nonumber \\ &
\Delta'_4 =  \alpha_{\frac{1}{2}, -\frac{1}{2},1}\alpha_{-\frac{1}{2}, -\frac{1}{2},0} - \alpha_{\frac{1}{2}, -\frac{1}{2},0} \alpha_{-\frac{1}{2}, -\frac{1}{2}, 1}~,~
\nonumber \\ &
\Delta_5 =  \alpha_{\frac{1}{2}, \frac{1}{2},1}\alpha_{-\frac{1}{2}, \frac{1}{2},-1} - \alpha_{\frac{1}{2}, \frac{1}{2},-1} \alpha_{-\frac{1}{2}, \frac{1}{2}, 1}~,~
\nonumber \\ &
\Delta'_5 =  \alpha_{\frac{1}{2}, -\frac{1}{2},1}\alpha_{-\frac{1}{2}, -\frac{1}{2},-1} - \alpha_{\frac{1}{2}, -\frac{1}{2},-1} \alpha_{-\frac{1}{2}, -\frac{1}{2}, 1}~,~
\nonumber \\ &
\Delta_6 =  \alpha_{\frac{1}{2}, \frac{1}{2},0}\alpha_{-\frac{1}{2}, \frac{1}{2},-1} - \alpha_{\frac{1}{2}, \frac{1}{2},-1} \alpha_{-\frac{1}{2}, \frac{1}{2}, 0}~,~
\nonumber \\ &
\Delta'_6 =  \alpha_{\frac{1}{2}, -\frac{1}{2},0}\alpha_{-\frac{1}{2}, -\frac{1}{2},-1} - \alpha_{\frac{1}{2}, -\frac{1}{2},-1} \alpha_{-\frac{1}{2}, -\frac{1}{2}, 0}~,~
\nonumber \\ &
\Delta_7 =  \alpha_{\frac{1}{2}, \frac{1}{2},1}\alpha_{-\frac{1}{2}, -\frac{1}{2},1} - \alpha_{\frac{1}{2}, -\frac{1}{2},1} \alpha_{-\frac{1}{2}, \frac{1}{2}, 1}~,~
\nonumber \\ &
\Delta'_7 =  \alpha_{\frac{1}{2}, \frac{1}{2},0}\alpha_{-\frac{1}{2}, -\frac{1}{2},0} - \alpha_{\frac{1}{2}, -\frac{1}{2},0} \alpha_{-\frac{1}{2}, \frac{1}{2}, 0}~,~
\nonumber \\ &
\Delta^{\prime \prime}_7 =  \alpha_{\frac{1}{2}, \frac{1}{2},-1}\alpha_{-\frac{1}{2}, -\frac{1}{2},-1} - \alpha_{\frac{1}{2}, -\frac{1}{2},-1} \alpha_{-\frac{1}{2}, \frac{1}{2}, -1}~,~
\nonumber \\ &
\Delta_8 =  \alpha_{\frac{1}{2}, \frac{1}{2},1}\alpha_{-\frac{1}{2}, -\frac{1}{2},0} - \alpha_{\frac{1}{2}, \frac{1}{2},0} \alpha_{-\frac{1}{2}, -\frac{1}{2}, 1}~,~
\nonumber \\ &
\Delta'_8 =  \alpha_{\frac{1}{2}, \frac{1}{2},1}\alpha_{-\frac{1}{2}, -\frac{1}{2},0} - \alpha_{\frac{1}{2}, -\frac{1}{2},1} \alpha_{-\frac{1}{2}, \frac{1}{2}, 0}~,~
\nonumber \\ &
\Delta^{\prime \prime}_8 =  \alpha_{\frac{1}{2}, \frac{1}{2},1}\alpha_{-\frac{1}{2}, -\frac{1}{2},0} - \alpha_{-\frac{1}{2}, \frac{1}{2},1} \alpha_{\frac{1}{2}, -\frac{1}{2}, 0}~,~
\nonumber \\ &
\Delta_9 =  \alpha_{\frac{1}{2}, \frac{1}{2},1}\alpha_{-\frac{1}{2}, -\frac{1}{2},-1} - \alpha_{\frac{1}{2}, \frac{1}{2},-1} \alpha_{-\frac{1}{2}, -\frac{1}{2}, 1}~,~
\nonumber \\ &
\Delta'_9 =  \alpha_{\frac{1}{2}, \frac{1}{2},1}\alpha_{-\frac{1}{2}, -\frac{1}{2},-1} - \alpha_{\frac{1}{2}, -\frac{1}{2},1} \alpha_{-\frac{1}{2}, \frac{1}{2}, -1}~,~
\nonumber \\ &
\Delta^{\prime \prime}_9 =  \alpha_{\frac{1}{2}, \frac{1}{2},1}\alpha_{-\frac{1}{2}, -\frac{1}{2},-1} - \alpha_{-\frac{1}{2}, \frac{1}{2},1} \alpha_{\frac{1}{2}, -\frac{1}{2}, -1}~,~
\nonumber \\ &
\Delta_{10} =  \alpha_{\frac{1}{2}, \frac{1}{2},0}\alpha_{-\frac{1}{2}, -\frac{1}{2},-1} - \alpha_{\frac{1}{2}, \frac{1}{2},-1} \alpha_{-\frac{1}{2}, -\frac{1}{2}, 0}~,~
\nonumber \\ &
\Delta'_{10}=  \alpha_{\frac{1}{2}, \frac{1}{2},0}\alpha_{-\frac{1}{2}, -\frac{1}{2},-1} - \alpha_{\frac{1}{2}, -\frac{1}{2},0} \alpha_{-\frac{1}{2}, \frac{1}{2}, -1}~,~
\nonumber \\ &
\Delta^{\prime \prime}_{10} =  \alpha_{\frac{1}{2}, \frac{1}{2},0}\alpha_{-\frac{1}{2}, -\frac{1}{2},-1} - \alpha_{-\frac{1}{2}, \frac{1}{2},0} \alpha_{\frac{1}{2}, -\frac{1}{2}, -1}~.~\,
\end{align}  
Thus, all the discriminants $\Delta_i$ are degree 2 homogeneous and holomorphic functions. 
Also, we can prove that the ranges of $\left|\Delta_i\right|$ are
\begin{align} 
 0 \leq \left|\Delta_i\right| \leq \frac{1}{2}~.~
\end{align}
Similar to the above discussions, the seven independent discriminants can be chosen as 
$\{\Delta_1, \Delta_2, \Delta_4, \Delta_5, \Delta_7, \Delta_8, \Delta_9\}$.
The classical space is the intersection of the discriminant loci $\Delta_i=0$ in quantum space, and the quantum entanglement space is the quantum space without
the intersection of the discriminant loci $\Delta_i=0$, {\it i.e.}, the quantum space minus the classical space.

\subsection{The General Systems with Two Particles and Three Particles }

We would like to study the discriminants for the two-particle system and three-particle system in general. We define the number of possible combinations of $m$ objects from a set of $n$ objects as $C(n,m)$
\begin{align}  
    C(n, m) = {{n(n-1)\cdot \cdot \cdot (n-m+1)}\over {m(m-1) \cdot \cdot \cdot 1}}~.~
\end{align}  

For a two-particle $P_1 P_2$ system with spin $s_1$ and $s_2$,
the most general polarization state can be written as 
\begin{align}  
    |P_1 P_2 \rangle = \sum^{s_1}_{j=-s_1} \sum_{k=-s_2}^{s_2} \alpha_{j,k} |j\rangle_{P_1} \otimes |k\rangle_{P_2}
\end{align}  
with the normalization condition  
\begin{align}  
\sum^{s_1}_{j=-s_1} \sum_{k=-s_2}^{s_2} |\alpha_{j,k}|^2 = 1~.  
\end{align}  

In the two-particle $P_1 P_2$  system with spin $s_1$ and $s_2$, the quantum space is $\mathbb{CP}^{J-1}$ 
with $J=(2s_1+1) \times (2s_2+1)$, which has complex dimension $(2s_1+1) \times (2s_2+1)-1$.
And the classical space is $\mathbb{CP}^{2s_1}\otimes \mathbb{CP}^{2s_2}$ with complex dimension $2s_1+2s_2$.
Thus, there are $(2s_1+1) \times (2s_2+1)-1-2s_1-2s_2=4s_1s_2$
independent discriminants with complex dimension 1. We define the general discriminants as
\begin{align} 
\Delta_{ijkl} =  \alpha_{i,j} \alpha_{k,l}-
\alpha_{i, l} \alpha_{k,j}~.~
\end{align}  
where $i, k =-s_1, \cdot \cdot \cdot, s_1$, and $j, l =-s_2, \cdot \cdot \cdot, s_2$. 
Thus, all the discriminants $\Delta_{ijkl}$ are degree 2 homogeneous and holomorphic functions, and the Number
of General Discriminants (NGD) is 
$C(2s_1+1, 2) \times C(2s_2+1, 2)$, {\it i.e.},
${\rm NGD}= C(2s_1+1, 2) \times C(2s_2+1, 2)$.
Also, we can prove that the ranges of $\left|\Delta_{ijkl}\right|$ are
\begin{align} 
 0 \leq \left|\Delta_{ijkl}\right| \leq \frac{1}{2}~.~
\end{align}
In addition, the classical space is 
the intersection of the discriminant loci $\Delta_{ijkl}=0$ in quantum space, and the quantum entanglement space is the quantum space without
the intersection of the discriminant loci $\Delta_{ijkl}=0$, {\it i.e.}, the quantum space minus the classical space.

In the three-particles $P_1 P_2 P_3$  system with spin $s_1$, $s_2$ and $s_3$, the quantum space is $\mathbb{CP}^{J-1}$ 
with $J=(2s_1+1) \times (2s_2+1) \times (2s_3+1)$, which has complex dimension $J-1=(2s_1+1) \times (2s_2+1)\times (2s_3+1) -1$.
And the classical space is $\mathbb{CP}^{2s_1}\otimes \mathbb{CP}^{2s_2}\otimes \mathbb{CP}^{2s_3}$ with complex dimension $2s_1+2s_2+2s_3$.
Thus, there are $(2s_1+1) \times (2s_2+1)\times (2s_3+1)-1-2s_1-2s_2-2s_3=8s_1s_2s_3+4s_1s_2+4s_1s_3+4s_2s_3$
independent discriminants with complex dimension 1. 
The strategy to construct the discriminants of $N+1$ particles is that we fix the spin (helicity) of one particle and construct the corresponding discriminants of $N$ particles, thus, we have $N+1$ kinds. Next, we consider the new discriminants where all the particles have different spins (helicities).
We define the general discriminants $\Delta_{ijklmn}$ as
\begin{align} 
& \Delta^I_{ijkimn} =  \alpha_{i,j,k} \alpha_{i, m,n}-
\alpha_{i, j, n} \alpha_{i,m,k}~,~
\nonumber \\ &
\Delta^{II}_{ijkljn} =  \alpha_{i,j,k} \alpha_{l, j,n}-
\alpha_{i, j, n} \alpha_{l,j,k}~,~
\nonumber \\ &
\Delta^{III}_{ijklmk} =  \alpha_{i,j,k} \alpha_{l, m,k}-
\alpha_{i, m, k} \alpha_{l,j,k}~,~
\nonumber \\ &
\Delta^{IV}_{ijklmn} =\alpha_{i,j,k} \alpha_{l, m,n}-
\alpha_{i, j, n} \alpha_{l,m,k}~,~
\nonumber \\ &
\Delta^{IV\prime }_{ijklmn} =\alpha_{i,j,k} \alpha_{l, m,n}-
\alpha_{i, m, k} \alpha_{l,j,n}~,~
\nonumber \\ &
\Delta^{IV\prime\prime}_{ijklmn} =\alpha_{i,j,k} \alpha_{l, m,n}-\alpha_{l, j, k} \alpha_{i,m,n}~,~
\end{align}  
where $i, l =-s_1, \cdot \cdot \cdot, s_1$,  $j, m =-s_2, \cdot \cdot \cdot, s_2$,
and $k, n =-s_3, \cdot \cdot \cdot, s_3$. Also,
$i\not=l$, $j\not=m$, and $k\not=n$ 
for the fourth kind of the discriminants $\Delta^{IV}_{ijklmn}$, $\Delta^{IV \prime}_{ijklmn}$, and
$\Delta^{IV \prime \prime}_{ijklmn}$.  
Thus, all the discriminants $\Delta_{ijklmn}$ are degree 2 homogeneous and holomorphic functions. Also, the Number
of General Discriminants (NGD) is 
\begin{align} 
{\rm NGD}=& (2s_1+1)C(2s_2+1, 2) C(2s_3+1, 2) 
+ (2s_2+1)C(2s_1+1, 2) C(2s_3+1, 2) 
\nonumber \\ &
+ (2s_3+1)C(2s_1+1, 2)  C(2s_2+1, 2) 
\nonumber \\ &
+ 3C(2s_1+1, 2)  C(2s_2+1, 2) C(2s_3+1, 2) ~.~
\end{align}
Moreover, we can prove that the ranges of $\left|\Delta_{ijklmn}\right|$ are
\begin{align} 
 0 \leq \left|\Delta_{ijklmn}\right| \leq \frac{1}{2}~.~
\end{align}
In addition, the classical space is 
the intersection of the discriminant loci $ \Delta_{ijklmn} =0$ in quantum space, and the quantum entanglement space is the quantum space without
the intersection of the discriminant loci $\Delta_{ijklmn}=0$, {\it i.e.}, the quantum space minus the classical space.

For any system of \( N \) particles labeled \( P_1 P_2 \ldots P_N \), we select \( m \) distinct particles (\( m \geq 2 \)) denoted as \( P_{i_1} P_{i_2} \ldots P_{i_m} \), with indices satisfying
\begin{align}  
1 \leq i_1 < i_2 < \ldots < i_m \leq N~.  
\end{align}  
We define the reduced coefficients by
\begin{align}  
\tilde{\alpha}_{k_{i_1},k_{i_2},\ldots,k_{i_m}} = \alpha_{k_1,k_2,\ldots,k_N}~~\text{with}~~k_j = s_j~~\text{for}~~j \in \{1,2,\ldots,N\} \setminus \{i_1,i_2,\ldots,i_m\}~.  
\end{align}  
The independent discriminants for the \( P_1 P_2 \ldots P_N \) system are given by  
\begin{align}  
\Delta_{k_{i_1},k_{i_2},\ldots, k_{i_m}} &= \tilde{\alpha}_{k_{i_1},k_{i_2},\ldots,k_{i_{m-1}},k_{i_m}}\tilde{\alpha}_{k_{i_1}+1,k_{i_2}+1,\ldots,k_{i_{m-1}}+1,k_{i_m}+1} \nonumber\\  
&\quad - \tilde{\alpha}_{k_{i_1},k_{i_2},\ldots,k_{i_{m-1}},k_{i_m}+1}\tilde{\alpha}_{k_{i_1}+1,k_{i_2}+1,\ldots,k_{i_{m-1}}+1,k_{i_m}}~,  
\end{align}  
where the spin projections are constrained by
\begin{align}  
k_{i_n} \in \{-s_{i_n}, -s_{i_n}+1, \ldots, s_{i_n}-1\}~,\quad n=1,2,\ldots,m~.  
\end{align}
The ranges of $\left|\Delta_{k_{i_1},k_{i_2},\ldots, k_{i_m}}\right|$ are
\begin{align} 
 0 \leq \left|\Delta_{k_{i_1},k_{i_2},\ldots, k_{i_m}}\right| \leq \frac{1}{2}~.~
\end{align}
We then compute the NID as
\begin{align}  
    \text{NID}=&\sum_{m=2}^N~\sum_{1\le i_1<i_2<\ldots i_m \le N}  
    ~\sum_{k_{i_1}=-s_{i_1}}^{s_{i_1}-1}\sum_{k_{i_2}=-s_{i_2}}^{s_{i_2}-1}\ldots \sum_{k_{i_m}=-s_{i_m}}^{s_{i_m}-1}1 \\  
    =& \prod_{i=1}^N(2s_i+1)-1-\sum_{i=1}^N 2s_i~,  
\end{align}  
which agrees with the result given in Eq.~(\ref{nnid}).  

\section{Theoretical framework for Specific Approach} \label{sec:15}

The most general polarization state of an N-particle system $P_1 P_2\ldots P_N$ can be represented as  
\begin{align}  
    |P_1 P_2\ldots P_N\rangle=\sum_{k_1,k_2,\ldots,k_N} \alpha_{k_1,k_2,\ldots,k_N}  
    |k_1\rangle_{P_1}|k_2\rangle_{P_2}\ldots|k_N\rangle_{P_N}
\end{align}  
with the normalization condition  
\begin{align}  
    \sum_{k_1,k_2,\ldots,k_N} \left|\alpha_{k_1,k_2,\ldots,k_N}\right|^2=1~.  
\end{align}  
Since the existence of quantum entanglement (QE) is independent of the choice of reference frame, we adopt the center-of-mass (c.m.) frame of the $P_1 P_2\ldots P_N$ system for our analysis without loss of generality. In this frame, the spin projection quantum numbers $k_i$ ($i=1,2,\ldots,N$) of particles $P_i$ are defined along their respective momentum directions, which are denoted as $\hat{e}_i$.

\subsection{Decay amplitudes and phase space integration}\label{211}

For the decay processes $P_i \to f_{i,1} + f_{i,2} + \ldots$ ($i=1,2,\ldots,N$), we collectively denote the final-state particles from $P_i$ decay as $f_i \equiv (f_{i,1}, f_{i,2}, \ldots)$. The transition amplitude for the whole $P_1 P_2\ldots P_N$ system decaying through these channels is given by  
\begin{align}  
    \mathcal{M} &=\sum_{k_1,k_2,\ldots,k_N}\alpha_{k_1,k_2,\ldots,k_N}  
    \langle f_1|k_1\rangle_{P_1}\langle f_2|k_2\rangle_{P_2}\ldots\langle f_N|k_N\rangle_{P_N}~.  
\end{align}  
We define  
\begin{align}  
    \Gamma=\int d\pi_{f_1} \int d\pi_{f_2}\ldots \int d\pi_{f_N} \left| \mathcal{M} \right|^2~,\label{mt1}
\end{align}  
where $d\pi_{f_i}$ $(i=1,2,\ldots,N)$ correspond to the phase space volume elements for the decay products of particles $P_i$. We directly obtain
\begin{align}
\Gamma &=\sum_{k_1,k_2,\ldots, k_N,k^\prime_1,k^\prime_2,\ldots, k^\prime_N}\alpha_{k_1,k_2,\ldots, k_N}\alpha_{k^\prime_1,k^\prime_2,\ldots, k^\prime_N}^* \prod_{i=1}^N\int d\pi_{f_i} \langle f_i|k_i\rangle_{P_i} \langle f_i|k_i^\prime\rangle_{P_i}^* ~.
\end{align}

\begin{itemize}  
    \item The analysis employs a fixed laboratory reference frame defined by orthonormal basis vectors,
\begin{align}
    \hat{e}_x=(1,0,0)~,\quad \hat{e}_y=(0,1,0)~,\quad \hat{e}_z=(0,0,1)~.
\end{align}
All particle momenta in this frame are subsequently determined with respect to these orthogonal basis vectors.
    \item The Lorentz invariance of $\int d\pi_{f_i} \langle f_i|k_i\rangle_{P_i} \langle f_i|k_i^\prime\rangle_{P_i}^*$ ($i=1,2,\ldots,N$) permits their evaluations in the respective rest frames of particles $P_i$. Crucially, the Lorentz transformations connecting the $P_1 P_2\ldots P_N$ c.m. frame to the rest frames of $P_i$ preserve the momentum directions of the boosted particles $P_i$. Consequently, the spin projection quantum number associated with each particle remains invariant under the respective transformation.
    \item We select one daughter particle from each decay channel: $f_{i,1}$ from $f_i$ ($i=1,2,\ldots,N$). Working in the rest frames of $P_i$, we define the polar angles $\theta_{i}$ and azimuthal angles $\phi_{i}$ for momentum directions of $f_{i,1}$, using as the polar axis $\hat{e}_i$. We use the following azimuthal angle reference protocol:  \\
-\textbf{Construct orthogonal bases}:  
   Choose auxiliary axes $\hat{e}^\prime_i$ orthogonal to $\hat{e}_i$ \\
-\textbf{Define zero azimuth}: Align $\phi_{i}=0$ with $\hat{e}^\prime_i$  \\
-\textbf{Angular measurement}: $\phi_{i} \in [0,2\pi]$ increase following the right-handed coordinate systems about $\hat{e}_{i}$  
    \item For a two-particle system $P_1 P_2$, the axial relationship $\hat{e}_1 = -\hat{e}_2$ holds in the c.m. frame. For computational consistency in two-particle system analysis, we impose the auxiliary axis condition $\hat{e}^\prime_1 = \hat{e}^\prime_2$.  The angular correlation between daughter particles $f_{1,1}$ and $f_{2,1}$ manifests through the angle $\theta_{12}$ defined by  
\begin{align}  
\cos\theta_{12}=&(\sin\theta_{1}\cos\phi_{1},\sin\theta_{1}\sin\phi_{1},\cos\theta_{1})\cdot(\sin\left(\pi-\theta_{2}\right)\cos\left(-\phi_{2}\right),\sin\left(\pi-\theta_{2}\right)\sin\left(-\phi_{2}\right),\cos\left(\pi-\theta_{2}\right)) \nonumber\\  
    =& -\cos\theta_{1}\cos\theta_{2}+\sin\theta_{1}\sin\theta_{2}\cos\left(\phi_{1}+\phi_{2}\right)~.  \label{jiajiao}
\end{align}
\end{itemize}

Following the formalism in Ref.~\cite{Leader:2001nas}, the decay amplitudes in the respective rest frames of $P_i$ are expressed as  
\begin{align}  
& \langle f_i|k_i\rangle_{P_i}=\sqrt{\frac{2s_i+1}{4\pi}} e^{i(k_i-\tilde{\lambda}_{f_i})\phi_{i}}  
d_{k_i,\tilde{\lambda}_{f_i}}^{s_i}(\theta_{i})H_{i}(\lambda_{f_i})~,  \quad i=1,2,\ldots,N~,
\end{align}  
where $s_{i}$ denote the spin quantum numbers of particles $P_i$ (e.g., $1/2$ for fermions, $1$ for vector bosons). $k_i$ represent the spin projection quantum numbers along the momentum directions of $P_i$ in the $P_1 P_2\ldots P_N$ c.m. frame. $\lambda_{f_{i}}$ encode polarization configurations: $\lambda_{f_i}=(\lambda_{f_{i,1}},\lambda_{f_{i,2}},\ldots)$, where $\lambda_{f_{i,j}}$ are spin projection numbers of particles $f_{i,j}$ defined relative to directions of $(\theta_{i},\phi_{i})$. The helicity summation rules are defined as $\tilde{\lambda}_{f_i} = \sum_j \lambda_{f_{i,j}}$. Crucially, $H_i(\lambda_{f_i})$ remains independent of both the angular variables ($\theta_{i},\phi_{i}$) and the parent particle spin projections $k_i$.
The Wigner $d$-functions satisfy the normalization conditions:  
\begin{align}  
    &\int_{-1}^1 d\cos\theta_{i} \left(d^{s_i}_{k_i,\tilde{\lambda}_{f_i}}(\theta_{i})\right)^2=\frac{2}{2s_i+1}~, \quad i=1,2,\ldots,N~.  
\end{align}

Through direct calculation, we derive the critical overlap integrals:  
\begin{align}
    &\int d\pi_{f_i} \langle f_i|k_i\rangle_{P_i} \langle f_i|k_i^\prime\rangle_{P_i}^* = \delta_{k_i}^{k_i^\prime} \int d\pi_{f_i}^\prime \left|H_i(\lambda_{f_i})\right|^2~, \quad i=1,2,\ldots,N,\label{mix}
\end{align}  
where the reduced phase space measure satisfies  
\begin{align}
    d\pi_{f_i}=d\pi_{f_i}^\prime d\phi_{i} d\cos\theta_{i}~.
\end{align}  
To obtain Eq.~(\ref{mix}), the angular integration exploits the orthogonality relation for integer $(k_i-k_i^\prime)$:  
\begin{align}
    \int_0^{2\pi}d\phi_{i} e^{i\left(k_i-k_i^\prime\right)\phi_{i}}=2\pi\delta_{k_i}^{k_i^\prime}~.
\end{align}
Then,  $\Gamma$ in Eq.~(\ref{mt1}) consequently simplifies to  
\begin{align}
    \Gamma &=\prod_{i=1}^N\int d\pi_{f_i}^\prime \left|H_i(\lambda_{f_i})\right|^2~.
\end{align}  
This result demonstrates that $\Gamma$ remains independent not only of the polarization coefficients $\alpha_{k_1,k_2,\ldots,k_N}$, but also of the momenta of particles $P_i$.

\subsection{Analytic structure of angular correlations in decay products}\label{for}

Given that $\theta_{i}$ and $\phi_{i}$ $(i=1,2,\ldots,N)$ represent measurable quantities, they naturally serve as building blocks for constructing composite observables $\mathcal{O}(\theta_{1},\theta_{2},\ldots,\theta_{N},\phi_{1},\phi_{2},\ldots,\phi_{N})$. The expectation value of such observables takes the general form:  
\begin{align}  
    &\langle \mathcal{O}(\theta_1,\theta_2,\ldots,\theta_N,\phi_1,\phi_2,\ldots,\phi_N)\rangle\nonumber\\
    =&
    \frac{\sum_{\lambda_{f_1},\lambda_{f_2},\ldots,\lambda_{f_N}} \int d\pi_{f_1} \int d\pi_{f_2}\ldots \int d\pi_{f_N} \mathcal{O}(\theta_1,\theta_2,\ldots,\theta_N,\phi_1,\phi_2,\ldots,\phi_N) \left| \mathcal{M} \right|^2}{\sum_{\lambda_{f_1},\lambda_{f_2},\ldots,\lambda_{f_N}} \Gamma}\nonumber\\
    =&\sum_{k_1,k_2,\ldots, k_N,k^\prime_1,k^\prime_2,\ldots, k^\prime_N}\mathcal{O}_{k_1,k_2,\ldots, k_N;k^\prime_1,k^\prime_2,\ldots, k^\prime_N}\alpha_{k_1,k_2,\ldots, k_N}\alpha^*_{k^\prime_1,k^\prime_2,\ldots, k^\prime_N}~.\label{msms}
\end{align} 
Direct calculation gives
\begin{align}  
    \mathcal{O}_{k_1,k_2,\ldots, k_N;k^\prime_1,k^\prime_2,\ldots, k^\prime_N}=& \prod_{i=1}^N \left( \frac{1}{\Gamma_{i}} \int d\pi_{f_{i}} e^{i\left(k_i-k^\prime_i\right)\phi_{i}}\sum_{\lambda_{f_{i}}}\left|H_{i}(\lambda_{f_{i}})\right|^2 d^{s_{i}}_{k_i,\tilde{\lambda}_{f_{i}}}(\theta_{i})d^{s_{i}}_{k^\prime_i,\tilde{\lambda}_{f_{i}}}(\theta_{i}) \right)\times\nonumber\\
    &\mathcal{O}(\theta_1,\theta_2,\ldots,\theta_N,\phi_1,\phi_2,\ldots,\phi_N)~, 
\end{align}  
where
\begin{align}  
  &\Gamma_{i}=\frac{4\pi}{2s_{i}+1}\int d\pi_{f_{i}}^\prime \left(\sum_{\lambda_{f_{i}}}\left|H_{i}(\lambda_{f_{i}})\right|^2\right)~,\quad i=1,2,\ldots,N~.  
\end{align}
When the observable admits the factorization:
\begin{align}
    \mathcal{O}(\theta_{1},\theta_{2},\ldots,\theta_{N},\phi_{1},\phi_{2},\ldots,\phi_{N})=\prod_{i=1}^N
\mathcal{O}^i(\theta_{i},\phi_{i})~,
\end{align}
the corresponding coefficients decompose as
\begin{align}  
    \mathcal{O}_{k_1,k_2,\ldots, k_N;k^\prime_1,k^\prime_2,\ldots, k^\prime_N}=\prod_{i=1}^N \mathcal{O}^i_{k_i,k_i^\prime}  
\end{align}  
with
\begin{align}  
    & \mathcal{O}^i_{k_i,k^\prime_i}= \frac{1}{\Gamma_{i}} \int d\pi_{f_{i}} \mathcal{O}^i(\theta_{i},\phi_{i}) e^{i\left(k_i-k^\prime_i\right)\phi_{i}}\sum_{\lambda_{f_{i}}}\left|H_{i}(\lambda_{f_{i}})\right|^2 d^{s_{i}}_{k_i,\tilde{\lambda}_{f_{i}}}(\theta_{i})d^{s_{i}}_{k^\prime_i,\tilde{\lambda}_{f_{i}}}(\theta_{i})~.
\end{align}  

The cosine-modulated observable reveals quantum correlations:  
\begin{align}  
    &\left\langle \left(\prod_{i=1}^N f^{i}(\theta_{i})\right)\cos\left(\sum_{i=1}^N d_i \phi_{i}\right)\right\rangle\nonumber\\
    =&\frac{1}{2} \sum_{k_1,k_2,\ldots,k_N}\left(\prod_{i=1}^N \mathcal{O}^{i}_{k_i,k_i+d_i}\right)\left(\alpha_{k_1,k_2,\ldots,k_N}\alpha_{k_1+d_1,k_2+d_2,\ldots,k_N+d_N}^*+\alpha_{k_1+d_1,k_2+d_2,\ldots,k_N+d_N}\alpha_{k_1,k_2,\ldots,k_N}^*\right) 
\end{align}
with
\begin{align}  
    & \mathcal{O}^{i}_{k_i,k_i+d_i}=\frac{1}{\Gamma_{i}}
        \left( \int d\pi_{f_{i}} f^{i}(\theta_{i}) 
  \left( \sum_{\lambda_{f_{i}}}\left|H_{i}(\lambda_{f_{i}})\right|^2   
     d^{s_{i}}_{k_i,\tilde{\lambda}_{f_{i}}}(\theta_{i}) d^{s_{i}}_{k_i+d_i,\tilde{\lambda}_{f_{i}}}(\theta_{i}) \right)\right)~.
\end{align} 
Similarly, we obtain
\begin{align}  
    &\left\langle \left(\prod_{i=1}^N f^{i}(\theta_{i})\right)\sin\left(\sum_{i=1}^N d_i \phi_{i}\right)\right\rangle\nonumber\\
    =&\frac{1}{2i} \sum_{k_1,k_2,\ldots,k_N}\left(\prod_{i=1}^N \mathcal{O}^{i}_{k_i,k_i+d_i}\right)\left(\alpha_{k_1,k_2,\ldots,k_N}\alpha_{k_1+d_1,k_2+d_2,\ldots,k_N+d_N}^*-\alpha_{k_1+d_1,k_2+d_2,\ldots,k_N+d_N}\alpha_{k_1,k_2,\ldots,k_N}^*\right) ~.
\end{align}

Considering the angular correlation $\cos\theta_{12}$ for the two-particle system in Eq.~(\ref{jiajiao}), the expectation value $\langle\cos\theta_{12}\rangle$ contains:  \\
- Diagonal terms $\left|\alpha_{k_1,k_2}\right|^2$ from the first term in the final line of Eq.~(\ref{jiajiao})   \\
- Coherence terms $\alpha_{k_1,k_2}\alpha^*_{k_1+1,k_2+1}+\alpha_{k_1+1,k_2+1}\alpha^*_{k_1,k_2}$ from the second term in the final line of Eq.~(\ref{jiajiao})   

\section{Revisiting quantum entanglement in $t\bar{t}$} \label{sec:2}

\subsection{The polarization state of $t\bar{t}$}

To accurately capture the polarization state of the $t\bar{t}$ system, which may exist in a superposition of different polarization states, its most complete representation is given by 
\begin{align}
    |t\bar{t}\rangle=\sum_{k,j=\pm \frac{1}{2}} \alpha_{k,j}|k\rangle_t |j\rangle_{\bar{t}}~,\label{tt1}
\end{align}
where $k$ and $j$ denote the helicities of $t$ and $\bar{t}$  along their respective momentum direction in the c.m. frame. For a given production channel, described as $initial~states \to t \bar{t} $, we define the amplitude of producing the $t\bar{t}$ pair in the state $|k\rangle_t |j\rangle_{\bar{t}}$ as $\tilde{\mathcal{M}}_{k,j}$. The coefficients $\alpha_{k,j}$ in Eq.~(\ref{tt1}) can be calculated using the expression
\begin{align}
    \alpha_{k,j}=\tilde{\mathcal{M}}_{k,j}/\sqrt{\sum_{k,j=\pm \frac{1}{2}} |\tilde{\mathcal{M}}_{k,j}|^2}~,\label{min}
\end{align}
which ensures that the normalization condition
\begin{align}
    \sum_{k,j=\pm \frac{1}{2}} |\alpha_{k,j}|^2=1\label{ai}
\end{align}
is satisfied.

It is important to unambiguously define the helicity state $|k\rangle_t |j\rangle_{\bar{t}}$. We adopt the same definitions for the polarizations of particles (including fermions, massive gauge bosons, and massless gauge bosons) as those presented in Appendices A.1 and A.2 of Ref.~\cite{Murayama:1992gi}. Accordingly, the helicity-eigenspinors for $|k\rangle_t$ and $|j\rangle_{\bar{t}}$, which align with the definitions of fermion polarizations in Ref.~\cite{Murayama:1992gi}, are given by
\begin{align}
    |\frac{1}{2}\rangle_t = \begin{pmatrix} 1 \\ 0 \end{pmatrix}, \quad 
|-\frac{1}{2}\rangle_t = \begin{pmatrix} 0 \\ 1 \end{pmatrix}, \quad 
|\frac{1}{2}\rangle_{\bar{t}} = \begin{pmatrix} 0 \\ -1 \end{pmatrix}, \quad 
|-\frac{1}{2}\rangle_{\bar{t}} = \begin{pmatrix} 1 \\ 0 \end{pmatrix}.
\end{align}
This allows us to represent the $t\bar{t}$ state as
\begin{align}
    |t\bar{t}\rangle=\sum_{k,j=\pm \frac{1}{2}} \alpha_{k,j}|k\rangle_t |j\rangle_{\bar{t}}=
    \begin{pmatrix} \alpha_{\frac{1}{2},-\frac{1}{2}} \\ -\alpha_{\frac{1}{2},\frac{1}{2}} \\ \alpha_{-\frac{1}{2},-\frac{1}{2}} \\ -\alpha_{-\frac{1}{2},\frac{1}{2}} \end{pmatrix}.
\end{align}
Consequently, the density matrix $\rho$ for the $t\bar{t}$ system is given by
\begin{align}
    \rho=|t\bar{t}\rangle \langle t\bar{t}|=
    \begin{pmatrix}
    |\alpha_{\frac{1}{2},-\frac{1}{2}}|^2 
    & -\alpha_{\frac{1}{2},-\frac{1}{2}}\alpha^*_{\frac{1}{2},\frac{1}{2}} 
    & \alpha_{\frac{1}{2},-\frac{1}{2}}\alpha^*_{-\frac{1}{2},-\frac{1}{2}} 
    & -\alpha_{\frac{1}{2},-\frac{1}{2}}\alpha^*_{-\frac{1}{2},\frac{1}{2}}  \\ 
     -\alpha_{\frac{1}{2},\frac{1}{2}}\alpha^*_{\frac{1}{2},-\frac{1}{2}} 
    & |\alpha_{\frac{1}{2},\frac{1}{2}}|^2 
    & -\alpha_{\frac{1}{2},\frac{1}{2}}\alpha^*_{-\frac{1}{2},-\frac{1}{2}} 
    & \alpha_{\frac{1}{2},\frac{1}{2}}\alpha^*_{-\frac{1}{2},\frac{1}{2}}  \\
    \alpha_{-\frac{1}{2},-\frac{1}{2}}\alpha^*_{\frac{1}{2},-\frac{1}{2}} 
    & -\alpha_{-\frac{1}{2},-\frac{1}{2}}\alpha^*_{\frac{1}{2},\frac{1}{2}} 
    & |\alpha_{-\frac{1}{2},-\frac{1}{2}}|^2
    & -\alpha_{-\frac{1}{2},-\frac{1}{2}}\alpha^*_{-\frac{1}{2},\frac{1}{2}}  \\
    -\alpha_{-\frac{1}{2},\frac{1}{2}}\alpha^*_{\frac{1}{2},-\frac{1}{2}} 
    & \alpha_{-\frac{1}{2},\frac{1}{2}}\alpha^*_{\frac{1}{2},\frac{1}{2}} 
    & -\alpha_{-\frac{1}{2},\frac{1}{2}}\alpha^*_{-\frac{1}{2},-\frac{1}{2}}
    & |\alpha_{-\frac{1}{2},\frac{1}{2}}|^2
    \end{pmatrix}.\label{rho2}
\end{align}
Since Eqs. (\ref{rho1}) and (\ref{rho2}) must yield the same density matrix $\rho$, we can straightforwardly derive the following relationships
\begin{align}
   & \rho_{11}+\rho_{44}=\frac{1}{2}(1+C_{33})=|\alpha_{\frac{1}{2},-\frac{1}{2}}|^2+|\alpha_{-\frac{1}{2},\frac{1}{2}}|^2~,\\
   & \rho_{23}+\rho_{32}=\frac{1}{2}(C_{11}+C_{22})=-\alpha_{\frac{1}{2},\frac{1}{2}}\alpha^*_{-\frac{1}{2},-\frac{1}{2}} -\alpha_{-\frac{1}{2},-\frac{1}{2}}\alpha^*_{\frac{1}{2},\frac{1}{2}}~.
\end{align}
From these results, we can express $D$ as
\begin{align}
   & D =\frac{1}{3} (C_{11}+C_{22}+C_{33})=\frac{2}{3}\left(|\alpha_{\frac{1}{2},-\frac{1}{2}}|^2+|\alpha_{-\frac{1}{2},\frac{1}{2}}|^2-\alpha_{\frac{1}{2},\frac{1}{2}}\alpha^*_{-\frac{1}{2},-\frac{1}{2}} -\alpha_{-\frac{1}{2},-\frac{1}{2}}\alpha^*_{\frac{1}{2},\frac{1}{2}}\right)-\frac{1}{3}~.\label{d1}
\end{align}

For the subsequent decay processes of $t \to  e^+ + \nu_e + b$ and $\bar{t} \to  e^- + \bar{\nu}_e + \bar{b}$ after the production of on-shell $t$ and $\bar{t}$, according to Ref.~\cite{Afik:2020onf,ATLAS2024,CMS:2024pts}, we have
\begin{align}
    D=-3\cdot\langle \cos \theta_{e^+ e^-} \rangle~,
\end{align}
where $\theta_{e^+ e^-}$ is the angle between the directions of the final-state $e^+$ and $e^-$ in the rest frames of $t$ and $\bar{t}$, respectively. Employing the same angular variable definitions established in Eq.~(\ref{jiajiao}), we derive the angular correlation:  
\begin{align}  
\cos\theta_{e^+ e^-} = -\cos\theta_{e^+}\cos\theta_{e^-} + \sin\theta_{e^+}\sin\theta_{e^-}\cos\left(\phi_{e^+} + \phi_{e^-}\right).  \label{jiajiao2}
\end{align}  

Building upon the theoretical framework outlined in Section~\ref{for}, direct calculations yield 
\begin{align}
    &\langle \cos \theta_{e^+ e^-} \rangle=\frac{1}{9}\left( |\alpha_{\frac{1}{2},\frac{1}{2}}|^2+|\alpha_{-\frac{1}{2},-\frac{1}{2}}|^2-|\alpha_{\frac{1}{2},-\frac{1}{2}}|^2-|\alpha_{-\frac{1}{2},\frac{1}{2}}|^2+2\alpha_{\frac{1}{2},\frac{1}{2}}\alpha^*_{-\frac{1}{2},-\frac{1}{2}}+2\alpha_{-\frac{1}{2},-\frac{1}{2}}\alpha^*_{\frac{1}{2},\frac{1}{2}}\right),\label{jjgg}
\end{align}
which aligns perfectly with Eq.~(\ref{d1}) through the relation established in Eq.~(\ref{ai}). This agreement provides rigorous verification of both the theoretical framework and computational methodology employed. Furthermore, following our preceding analysis, the result in Eq.~({\ref{jjgg}}) can be decomposed into distinct angular correlation components: 
\begin{align}
    &\langle  -\cos\theta_{e^+}\cos\theta_{e^-} \rangle=\frac{1}{9}\left( |\alpha_{\frac{1}{2},\frac{1}{2}}|^2+|\alpha_{-\frac{1}{2},-\frac{1}{2}}|^2-|\alpha_{\frac{1}{2},-\frac{1}{2}}|^2-|\alpha_{-\frac{1}{2},\frac{1}{2}}|^2\right), \\
    &\langle  \sin\theta_{e^+}\sin\theta_{e^-}\cos\left(\phi_{e^+} + \phi_{e^-}\right) \rangle=\frac{2}{9}\left(\alpha_{\frac{1}{2},\frac{1}{2}}\alpha^*_{-\frac{1}{2},-\frac{1}{2}}+\alpha_{-\frac{1}{2},-\frac{1}{2}}\alpha^*_{\frac{1}{2},\frac{1}{2}}\right). \label{pi1}
\end{align}

\subsection{Criterion for quantum entanglement by $\cos \theta_{e^+ e^-}$}     

The criteria of $D < -\frac{1}{3}$ is a Sufficient but Not Necessary Condition for observing QE in the $t\bar{t}$ system as stated in Ref.~\cite{Afik:2020onf}. To enhance the understanding of the above, this section firstly presents a detailed derivation of the necessary and sufficient conditions for QE observation.

Assuming a state of the absence of QE between $t$ and $\bar{t}$, the polarization states of those two particles then become independent of each other and the $t\bar{t}$ system can be described by 
\begin{align}
    |t\bar{t}\rangle=\left(\sum_{k=\pm \frac{1}{2}}\beta_{k}|k\rangle_t\right)\left(\sum_{j=\pm \frac{1}{2}}\gamma_{j}|j\rangle_{\bar{t}}\right)~,\label{a1}
\end{align}
subject to the normalization conditions
\begin{align}
   \sum_{k=\pm \frac{1}{2}} |\beta_{k}|^2=\sum_{j=\pm \frac{1}{2}}|\gamma_{j}|^2=1~. \label{at}
\end{align}
Under these conditions, the coefficients $\alpha_{k,j}$ are given by
\begin{align}
    \alpha_{k,j}=\beta_k \gamma_j~,\quad k,j=\pm \frac{1}{2}~.\label{a2}
\end{align}
This formulation describes a state where the polarizations of $t$ and $\bar{t}$ are factorizable, indicating the absence of QE.
From Eq.~(\ref{a2}), we derive the following result
\begin{align}
    \alpha_{\frac{1}{2},\frac{1}{2}}\alpha_{-\frac{1}{2},-\frac{1}{2}}-\alpha_{\frac{1}{2},-\frac{1}{2}}\alpha_{-\frac{1}{2},\frac{1}{2}}=0~.\label{a3}
\end{align}
Moreover, if Eq.~(\ref{a3}) holds true, the state $|t\bar{t}\rangle$ can be represented in the form of Eq.~(\ref{a1}) with
\begin{align}
    &\beta_{\pm\frac{1}{2}}=\frac{\alpha_{\pm\frac{1}{2},\frac{1}{2}}}{\sqrt{|\alpha_{\frac{1}{2},\frac{1}{2}}|^2+|\alpha_{-\frac{1}{2},\frac{1}{2}}|^2}}~, \quad \gamma_{\pm\frac{1}{2}}=\frac{\alpha_{\frac{1}{2},\pm\frac{1}{2}}}{\sqrt{|\alpha_{\frac{1}{2},\frac{1}{2}}|^2+|\alpha_{\frac{1}{2},-\frac{1}{2}}|^2}}~,
\end{align}
where, without loss of generality, we assume $\alpha_{\frac{1}{2},\frac{1}{2}} = |\alpha_{\frac{1}{2},\frac{1}{2}}|$. Consequently, the condition
\begin{align}
   \Delta=\alpha_{\frac{1}{2},\frac{1}{2}}\alpha_{-\frac{1}{2},-\frac{1}{2}}-\alpha_{\frac{1}{2},-\frac{1}{2}}\alpha_{-\frac{1}{2},\frac{1}{2}}\ne0
\end{align}
is a Sufficient and Necessary Condition for indicating QE in the $t\bar{t}$ system.

With the above Sufficient and Necessary Condition, one can demonstrate the condition $D < -\frac{1}{3}$, where $D$ is defined in Eq.~(\ref{d1}), is a Sufficient but Not Necessary criterion for establishing QE in the $t\bar{t}$ system. The details are listed below. This chain of logic confirms that $D < -\frac{1}{3}$ suffices to indicate QE of $t\bar{t}$, without necessarily being the only indicator.
\begin{align}
     & D =\frac{2}{3}\left(|\alpha_{\frac{1}{2},-\frac{1}{2}}|^2+|\alpha_{-\frac{1}{2},\frac{1}{2}}|^2-\alpha_{\frac{1}{2},\frac{1}{2}}\alpha^*_{-\frac{1}{2},-\frac{1}{2}} -\alpha_{-\frac{1}{2},-\frac{1}{2}}\alpha^*_{\frac{1}{2},\frac{1}{2}}\right)-\frac{1}{3}<-\frac{1}{3} \nonumber\\
     & \Longrightarrow |\alpha_{\frac{1}{2},-\frac{1}{2}}|^2+|\alpha_{-\frac{1}{2},\frac{1}{2}}|^2<\alpha_{\frac{1}{2},\frac{1}{2}}\alpha^*_{-\frac{1}{2},-\frac{1}{2}} +\alpha_{-\frac{1}{2},-\frac{1}{2}}\alpha^*_{\frac{1}{2},\frac{1}{2}} \nonumber\\
     & \Longrightarrow 2|\alpha_{\frac{1}{2},-\frac{1}{2}}| |\alpha_{-\frac{1}{2},\frac{1}{2}}|<|\alpha_{\frac{1}{2},-\frac{1}{2}}|^2+|\alpha_{-\frac{1}{2},\frac{1}{2}}|^2<\alpha_{\frac{1}{2},\frac{1}{2}}\alpha^*_{-\frac{1}{2},-\frac{1}{2}} +\alpha_{-\frac{1}{2},-\frac{1}{2}}\alpha^*_{\frac{1}{2},\frac{1}{2}}<2|\alpha_{\frac{1}{2},\frac{1}{2}}| |\alpha_{-\frac{1}{2},-\frac{1}{2}}| \nonumber\\
      & \Longrightarrow |\alpha_{\frac{1}{2},-\frac{1}{2}}| |\alpha_{-\frac{1}{2},\frac{1}{2}}|<|\alpha_{\frac{1}{2},\frac{1}{2}}| |\alpha_{-\frac{1}{2},-\frac{1}{2}}| \nonumber\\
      & \Longrightarrow |\alpha_{\frac{1}{2},-\frac{1}{2}} \alpha_{-\frac{1}{2},\frac{1}{2}}|<|\alpha_{\frac{1}{2},\frac{1}{2}} \alpha_{-\frac{1}{2},-\frac{1}{2}}| \nonumber\\
       & \Longrightarrow |\alpha_{\frac{1}{2},\frac{1}{2}}\alpha_{-\frac{1}{2},-\frac{1}{2}}-\alpha_{\frac{1}{2},-\frac{1}{2}}\alpha_{-\frac{1}{2},\frac{1}{2}}|>0~.
\end{align}

A further investigation is to determine the range of possible values for $D$ based on whether Eq.~(\ref{a2}) is satisfied. When Eq.~(\ref{a2}) holds, we can, without loss of generality, use the following parameterization
\begin{align}
    \beta_{\frac{1}{2}}=\cos b_1~,\quad \beta_{-\frac{1}{2}}=\sin b_1~e^{i\eta_1}~,\quad \gamma_{\frac{1}{2}}=\cos b_2~,\quad \gamma_{-\frac{1}{2}}=\sin b_2~e^{i\eta_2}~.\label{b12}
\end{align}
By substituting Eqs.~(\ref{b12}) and (\ref{a2}) into Eq.~(\ref{d1}) and varying $b_1$, $b_2$, $\eta_1$, and $\eta_2$ within the interval $[0, 2\pi]$, we find that $D$ ranges as follows
\begin{align}
    D\in [-\frac{1}{3},\frac{1}{3}]~.\label{ran1}
\end{align}
In the most general case, where Eq.~(\ref{a2}) might not be satisfied, we can adopt this parametrization
\begin{align}
    \alpha_{\frac{1}{2},\frac{1}{2}}=\cos a_1\cos a_2~,~\alpha_{\frac{1}{2},-\frac{1}{2}}=\cos a_1\sin a_2~e^{i\xi_1}~,~ \alpha_{-\frac{1}{2},\frac{1}{2}}=\sin a_1 \cos a_3~e^{i\xi_2}~,~ \alpha_{-\frac{1}{2},-\frac{1}{2}}=\sin a_1 \sin a_3~e^{i\xi_3}~.\label{a12}
\end{align}
By inserting Eq.~(\ref{a12}) into Eq.~(\ref{d1}) and exploring $a_1$, $a_2$, $a_3$, $\xi_1$, $\xi_2$, and $\xi_3$ over the range $[0, 2\pi]$, we determine that $D$ spans the range
\begin{align}
    D\in [-1,\frac{1}{3}]~.\label{ran2}
\end{align}
Therefore, values of $D$ that lie within the range specified by Eq.~(\ref{ran2}) but outside the range given by Eq.~(\ref{ran1}), i.e.,
\begin{align}
    D\in [-1,-\frac{1}{3})~,
\end{align}
serve as a sufficient but not necessary condition for indicating QE in the $t\bar{t}$ system. 

\begin{figure}[tbp]
\begin{center}
\includegraphics[width=0.48\textwidth]{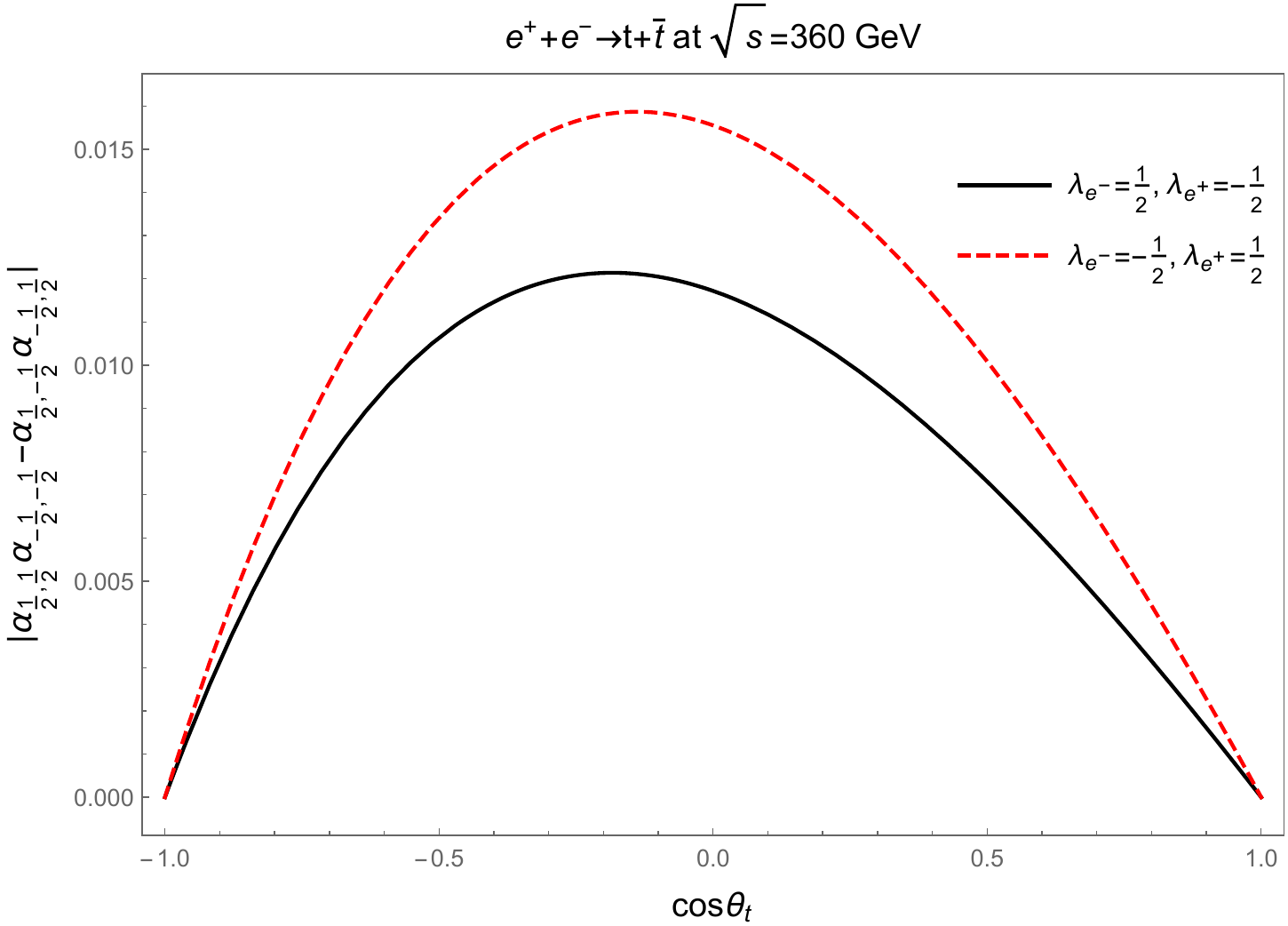} \hfill
\includegraphics[width=0.48\textwidth]{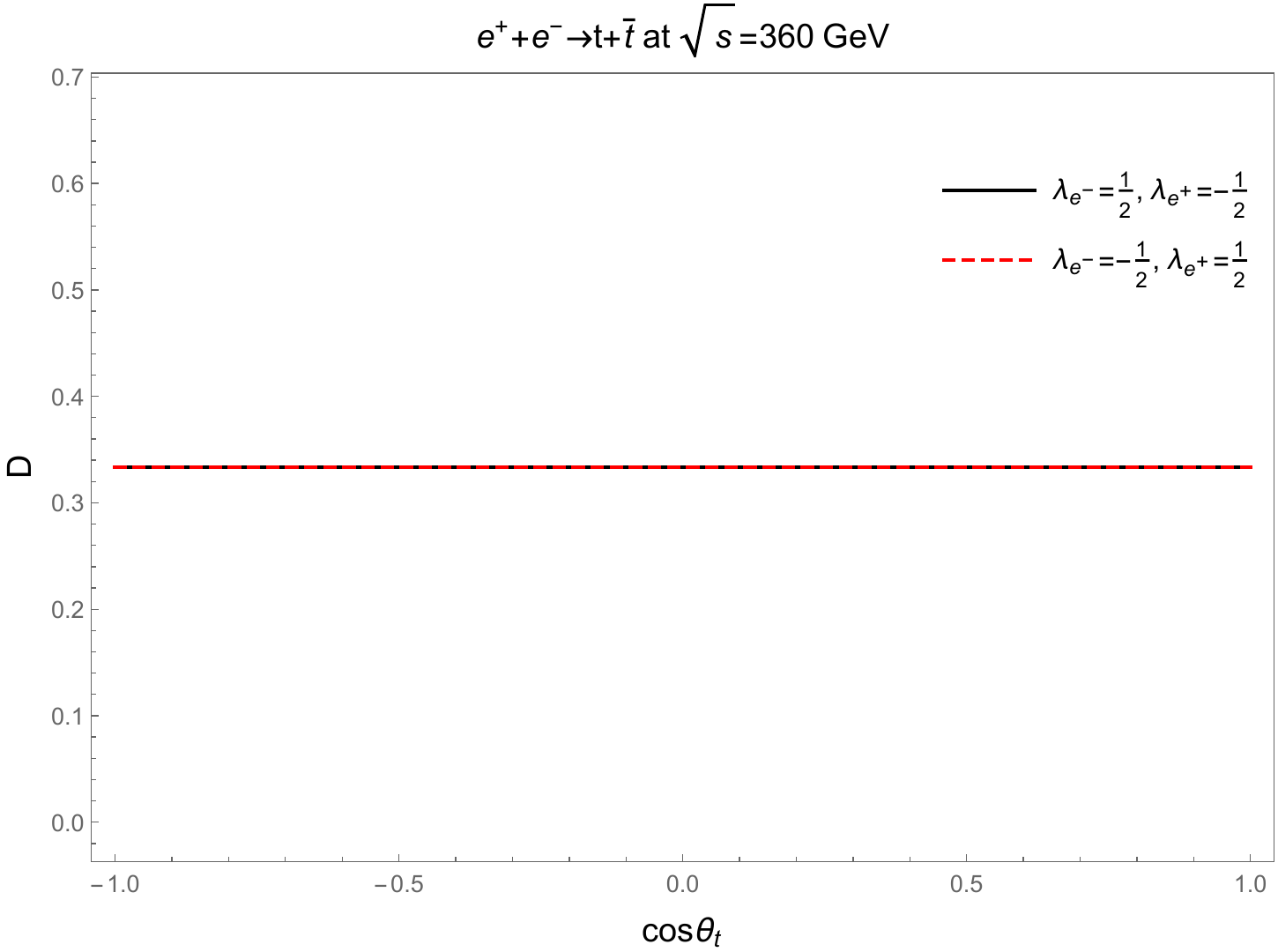}
\end{center}
\caption{
The LO predictions of $ |\alpha_{\frac{1}{2},\frac{1}{2}}\alpha_{-\frac{1}{2},-\frac{1}{2}} - \alpha_{\frac{1}{2},-\frac{1}{2}}\alpha_{-\frac{1}{2},\frac{1}{2}}| $ and $ D = -3 \cdot \langle \cos \theta_{e^+ e^-} \rangle $ for $ t\bar{t} $ pairs produced at an $ e^+ e^- $ collider operating at a c.m. energy of $\sqrt{s} = 360$ GeV. Here, $ \theta_t $ denotes the polar angle of the top quark $ t $ in the laboratory frame. The symbols $ \lambda_{e^\pm} $ represent the helicities of the $ e^+ $ and $ e^- $ beams, defined along their respective momentum directions in the laboratory frame.
\label{cy}}
\end{figure}

In Figure \ref{cy}, we display the LO predictions of $ |\alpha_{\frac{1}{2},\frac{1}{2}} \alpha_{-\frac{1}{2},-\frac{1}{2}} - \alpha_{\frac{1}{2},-\frac{1}{2}} \alpha_{-\frac{1}{2},\frac{1}{2}}| $ and $ D=-3\cdot\langle \cos \theta_{e^+ e^-} \rangle $ for $ t\bar{t} $ pairs produced at an $ e^+ e^- $ collider with a c.m. energy of $\sqrt{s}=360$ GeV, considering various polarization states of the $ e^+ $ and $ e^- $ beams. The condition $ |\alpha_{\frac{1}{2},\frac{1}{2}} \alpha_{-\frac{1}{2},-\frac{1}{2}} - \alpha_{\frac{1}{2},-\frac{1}{2}} \alpha_{-\frac{1}{2},\frac{1}{2}}| > 0 $ is both necessary and sufficient to demonstrate the presence of QE in the $ t\bar{t} $ system. From the left panel of Figure \ref{cy}, we conclude that QE exists in the $ t\bar{t} $ system if $\theta_t \neq 0$ or $\pi$. However, the LO results at the $ e^+ e^- $ collider satisfy the relation
\begin{align}
    &\alpha_{\frac{1}{2},\frac{1}{2}} = -\alpha_{-\frac{1}{2},-\frac{1}{2}}~,\label{tj1}\\
    &\left|\alpha_{\frac{1}{2},\frac{1}{2}}\right|^2 = \left|\alpha_{-\frac{1}{2},-\frac{1}{2}}\right|^2\le \frac{1}{4}~,\label{tj2}
\end{align}
which holds true for other beam energies as well, resulting in $ D = \frac{1}{3} $, as depicted in the right panel of Figure \ref{cy}. Therefore, the measurement of $ D $ at the $ e^+ e^- $ collider does not provide sufficient evidence for QE in the $ t\bar{t} $ system.

\subsection{Other observables and criteria}

However, guided by our master formula in Eq.~(\ref{msms}), we can construct alternative observables to probe QE in $t\bar{t}$ pairs produced at $e^+ e^-$ collider:

\begin{itemize}
\item The calculation at the LO level yields  
\begin{align}
    \langle \cos\left(\phi_{e^+}-\phi_{e^-}\right)\rangle = \frac{\pi^2}{32}\left(\alpha_{-\frac{1}{2},\frac{1}{2}}\alpha_{\frac{1}{2},-\frac{1}{2}}^* + \alpha_{\frac{1}{2},-\frac{1}{2}}\alpha_{-\frac{1}{2},\frac{1}{2}}^*\right).
\end{align}  
So, we define the normalized entanglement witness  
\begin{align}
    D^\prime = \frac{32}{\pi^2}\langle \cos\left(\phi_{e^+}-\phi_{e^-}\right)\rangle = \left(\alpha_{-\frac{1}{2},\frac{1}{2}}\alpha_{\frac{1}{2},-\frac{1}{2}}^* + \alpha_{\frac{1}{2},-\frac{1}{2}}\alpha_{-\frac{1}{2},\frac{1}{2}}^*\right),
\end{align}  
which fundamentally satisfies
\begin{align}
    D^\prime \in [-1, 1]~.
\end{align}  
Under the constraints of Eq.~(\ref{a2}), this observable is bounded by  
\begin{align}
    D^\prime \in \left[-\tfrac{1}{2}, \tfrac{1}{2}\right]~.
\end{align}  
Consequently, the existence of QE in $t\bar{t}$ pairs is conclusively demonstrated when  
\begin{align}
    D^\prime \in [-1, -\tfrac{1}{2}) \cup (\tfrac{1}{2}, 1]~.
\end{align}

\begin{figure}[tbp]
\begin{center}
\includegraphics[width=0.48\textwidth]{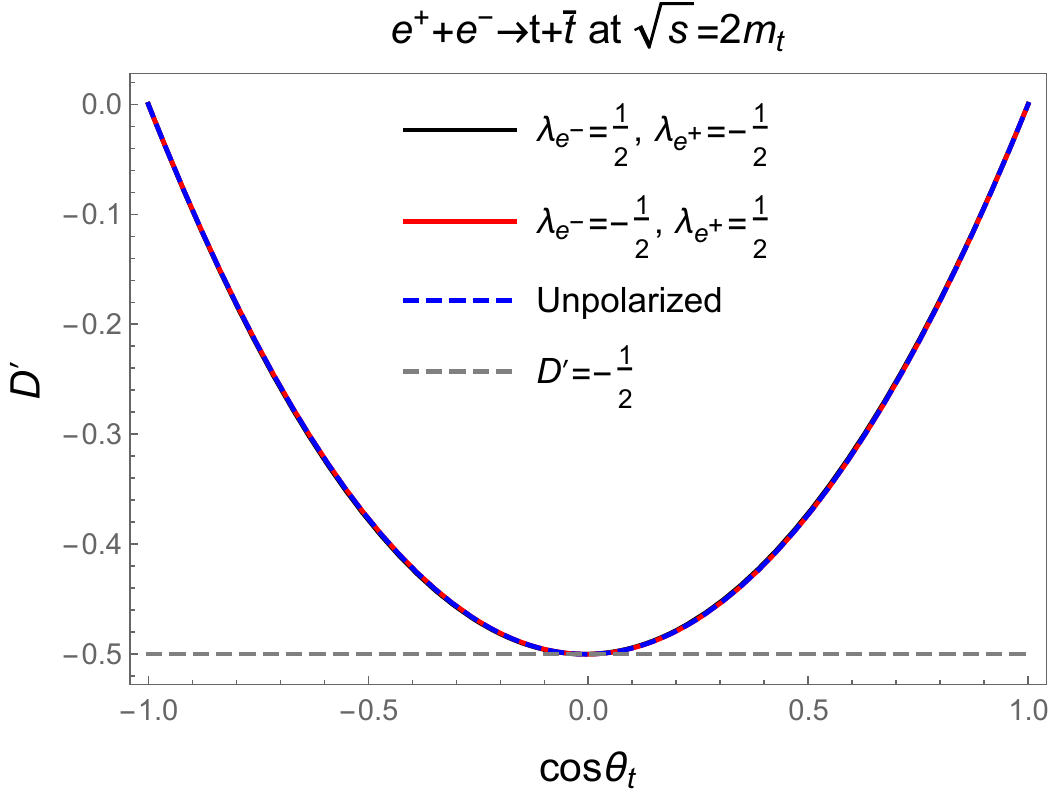} \\
\includegraphics[width=0.48\textwidth]{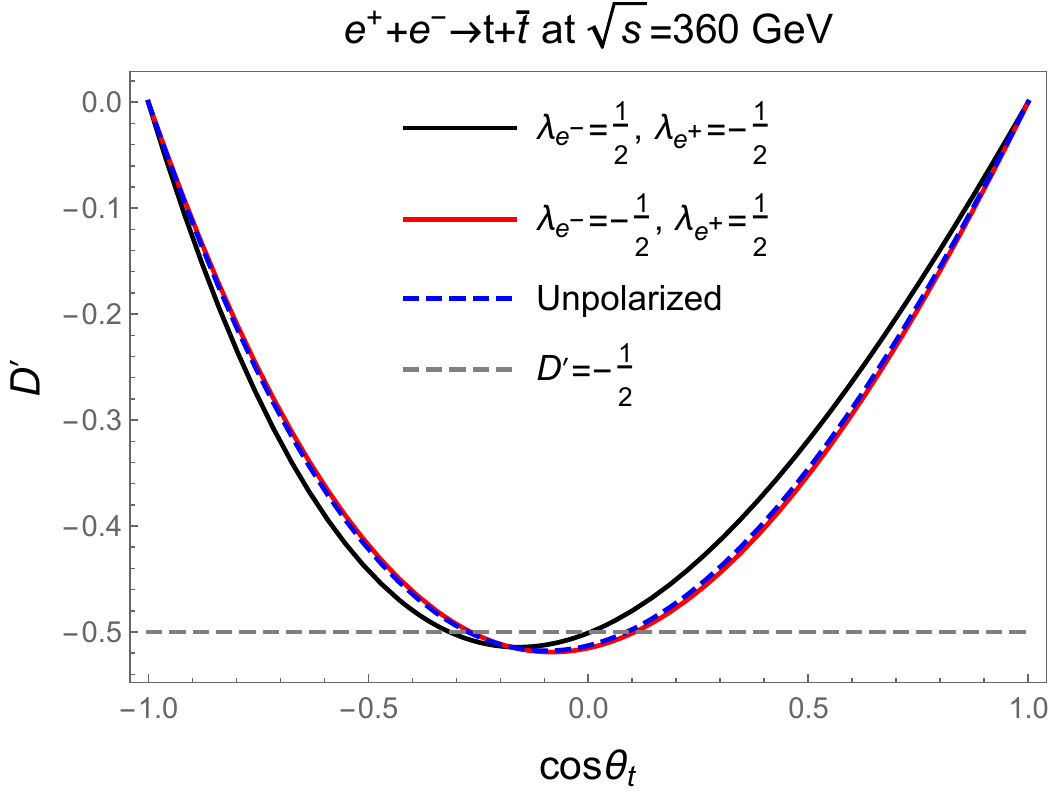}\hfill
\includegraphics[width=0.48\textwidth]{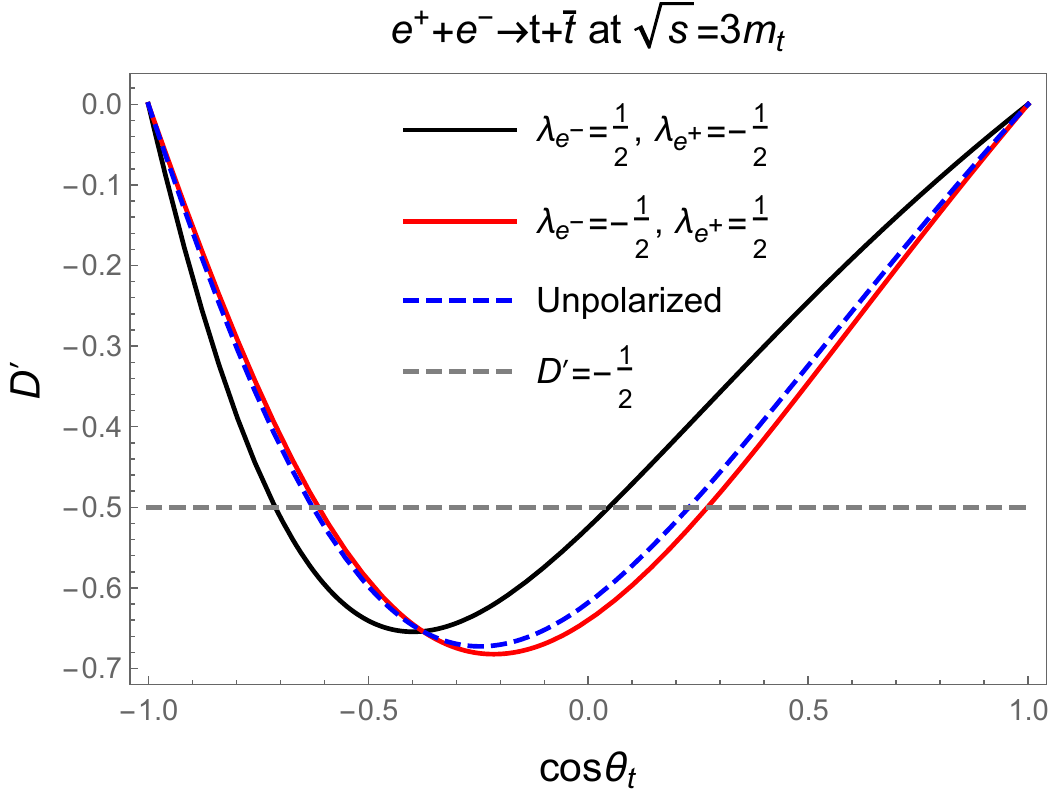}
\end{center}
\caption{The LO predictions of $ D^{\prime}=\frac{32}{\pi^2}\langle \cos\left(\phi_{e^+}-\phi_{e^-}\right)\rangle $ for $ t\bar{t} $ pairs produced at $ e^+ e^- $ collider with $\sqrt{s}=2m_t$, $360$ GeV, and $3m_t$, respectively. The angle $ \theta_t $ represents the polar angle of the top quark $ t $ in the laboratory frame. The symbols $ \lambda_{e^\pm} $ indicate the helicities of the $ e^+ $ and $ e^- $ beams, defined along their respective momentum directions in the laboratory frame.
\label{dp}}
\end{figure}

Figure \ref{dp} displays the LO predictions of $ D^{\prime}=\frac{32}{\pi^2}\langle \cos\left(\phi_{e^+}-\phi_{e^-}\right)\rangle $ derived from $t\bar{t}$ production at various $e^+ e^-$ collider energies, incorporating different initial-state polarization configurations. Our analysis reveals that above the $t\bar{t}$ production threshold ($2m_t$), specific angular regions of top-quark pair production (characterized by $\theta_t$ ranges) yield $D^\prime < -\frac{1}{2}$, thus serving as conclusive evidence for quantum entanglement in $t\bar{t}$ systems.

\item Motivated by Eq.~(\ref{pi1}), we define the azimuthal correlation observable: 
\begin{align}  
D^{\prime\prime} = \frac{9}{2}\langle  \sin\theta_{e^+}\sin\theta_{e^-}\cos\left(\phi_{e^+} + \phi_{e^-}\right) \rangle = \alpha_{\frac{1}{2},\frac{1}{2}}\alpha^*_{-\frac{1}{2},-\frac{1}{2}} + \alpha_{-\frac{1}{2},-\frac{1}{2}}\alpha^*_{\frac{1}{2},\frac{1}{2}}~. \label{pi}  
\end{align}  
Similarly, the condition of  
\begin{align}  
D^{\prime\prime} \in [-1, -\tfrac{1}{2}) \cup (\tfrac{1}{2}, 1]
\end{align}  
serves as a sufficient criterion for quantum entanglement in $t\bar{t}$ systems.  
Within the SM at $e^+ e^-$ colliders, using Eqs.~(\ref{tj1}) and (\ref{tj2}), we obtain  
\begin{align}  
D^{\prime\prime} \in [-\tfrac{1}{2}, 0)~.  
\end{align}  
This range cannot substantiate $t\bar{t}$ entanglement.  

However, considering beyond-Standard-Model scenarios with a Higgs-like particle exhibiting Yukawa coupling
\begin{align}  
\propto h^\prime t\bar{t}~.
\end{align}  
Direct calculation reveals that $t\bar{t}$ pairs from $h^\prime\to t+\bar{t}$ decays satisfy
\begin{align}  
\alpha_{\frac{1}{2},\frac{1}{2}} = -\alpha_{-\frac{1}{2},-\frac{1}{2}}~,\quad \left|\alpha_{\frac{1}{2},\frac{1}{2}}\right|^2 = \left|\alpha_{-\frac{1}{2},-\frac{1}{2}}\right|^2 = \frac{1}{2}~.  
\end{align}  
For such $h^\prime$-mediated $t\bar{t}$ production,  
\begin{align}  
D^{\prime\prime} = -1~,  
\end{align}  
demonstrating the utility of $D^{\prime\prime}$ as an entanglement witness in exotic decay channels.  

\item Through theoretical analysis of the $t\bar{t}$ system, we establish that the coefficients $B^\pm_k$ ($k=1,2,3$) and $C_{ij}$ ($i,j=1,2,3$) in Eqs.~(\ref{bb1})--(\ref{bbn}) correspond to statistical averages of measurable quantities as follows:  
\begin{align}  
   & B^\pm_k =3 \langle q^\pm_k \rangle~, \\  
   & C_{ij}=9 \langle q^+_i q^-_j \rangle~,  
\end{align}  
where the angular vectors are defined as  
\begin{align}  
   & q^+ = \left(\sin\theta_{e^+}\cos\phi_{e^+},\,\sin\theta_{e^+}\sin\phi_{e^+},\,\cos\theta_{e^+} \right)~, \\  
  & q^- = \left(\sin(\pi-\theta_{e^-})\cos\phi_{e^-},\,\sin(\pi-\theta_{e^-})\sin\phi_{e^-},\,\cos(\pi-\theta_{e^-}) \right)~.  
\end{align} 
The magnitude of quantum entanglement, parameterized by $|\Delta|$, can be experimentally determined through eigenvalue analysis of $C^\text{T}C$ using measured $C_{ij}$ values.  
When $\Delta=0$ (indicating no quantum entanglement in the $t\bar{t}$ system), the correlation matrix simplifies to:  
\begin{align}  
    C_{ij}=B^+_i B^-_j~.  
\end{align}  
\end{itemize}

\subsection{Similar case: quantum entanglement in $\tau^+\tau^-$ at $e^+ e^-$ collider}

As $\tau^+\tau^-$ system constitutes a fermion pair analogous to $t\bar{t}$ in $e^+ e^-$ collisions, we investigate their QE through the decay channels $\tau^+\to e^+ +\nu_e+\bar{\nu}_\tau$ and $\tau^-\to e^- +\bar{\nu}_e+\nu_\tau$. The LO computation yields the angle correlation:
\begin{align}
    \langle \cos \theta_{e^+ e^-} \rangle &=C_{\tau}^2\left( |\alpha_{\frac{1}{2},\frac{1}{2}}|^2+|\alpha_{-\frac{1}{2},-\frac{1}{2}}|^2-|\alpha_{\frac{1}{2},-\frac{1}{2}}|^2-|\alpha_{-\frac{1}{2},\frac{1}{2}}|^2+2\alpha_{\frac{1}{2},\frac{1}{2}}\alpha^*_{-\frac{1}{2},-\frac{1}{2}}+2\alpha_{-\frac{1}{2},-\frac{1}{2}}\alpha^*_{\frac{1}{2},\frac{1}{2}}\right)\\
    &=C_{\tau}^2\times (-1)
\end{align}
with
\begin{align}
    & C_{\tau} \nonumber\\
    =&\frac{1}{9}
    \frac{-24 r^2 \text{Li}_2\left(1-\frac{1}{r^2}\right)+\left(4 \pi ^2-6\right) r^2+\frac{1}{r^2}+6 \left(r^4+r^2-8 r^2 \ln (r)-2\right) \ln \left(\frac{r^2}{r^2-1}\right)-33}{2 r^2-\frac{1}{3 r^2}-2r^2 \left(r^2-1\right) \ln \left(\frac{r^2}{r^2-1}\right)-1}~, \\
    & r=\frac{m_W}{m_\tau}~,
\end{align}
where $\theta_{e^+ e^-}$ is the angle between the directions of the final-state $e^+$ and $e^-$ in the rest frames of $\tau^+$ and $\tau^-$, respectively, thus rendering it ineffective for demonstrating QE in $\tau^+ \tau^-$.

Besides, LO calculations yield the azimuthal angular correlation:
\begin{align}
   & \langle \cos\left(\phi_{e^+}-\phi_{e^-}\right)\rangle =C^\prime_{\tau^+ \tau^-}\left(\alpha_{-\frac{1}{2},\frac{1}{2}}\alpha_{\frac{1}{2},-\frac{1}{2}}^* + \alpha_{\frac{1}{2},-\frac{1}{2}}\alpha_{-\frac{1}{2},\frac{1}{2}}^*\right), \\
    & C^\prime_{\tau^+ \tau^-}=\frac{9}{32}\pi^2 C_{\tau}^2~.
\end{align}
We therefore define the normalized entanglement witness:
\begin{align}
    D^\prime_{\tau^+\tau^-} = \frac{1}{C^\prime_{\tau^+ \tau^-}}\langle \cos\left(\phi_{e^+}-\phi_{e^-}\right)\rangle = \left(\alpha_{-\frac{1}{2},\frac{1}{2}}\alpha_{\frac{1}{2},-\frac{1}{2}}^* + \alpha_{\frac{1}{2},-\frac{1}{2}}\alpha_{-\frac{1}{2},\frac{1}{2}}^*\right).
\end{align}  
So,
\begin{align}
   D^\prime_{\tau^+\tau^-} \in [-1, -\tfrac{1}{2}) \cup (\tfrac{1}{2}, 1]~.
\end{align}
serves as a sufficient criterion for demonstrating QE in $\tau^+\tau^-$ pairs produced at $e^+ e^-$ colliders.

\begin{figure}[tbp]
\begin{center}
\includegraphics[width=0.48\textwidth]{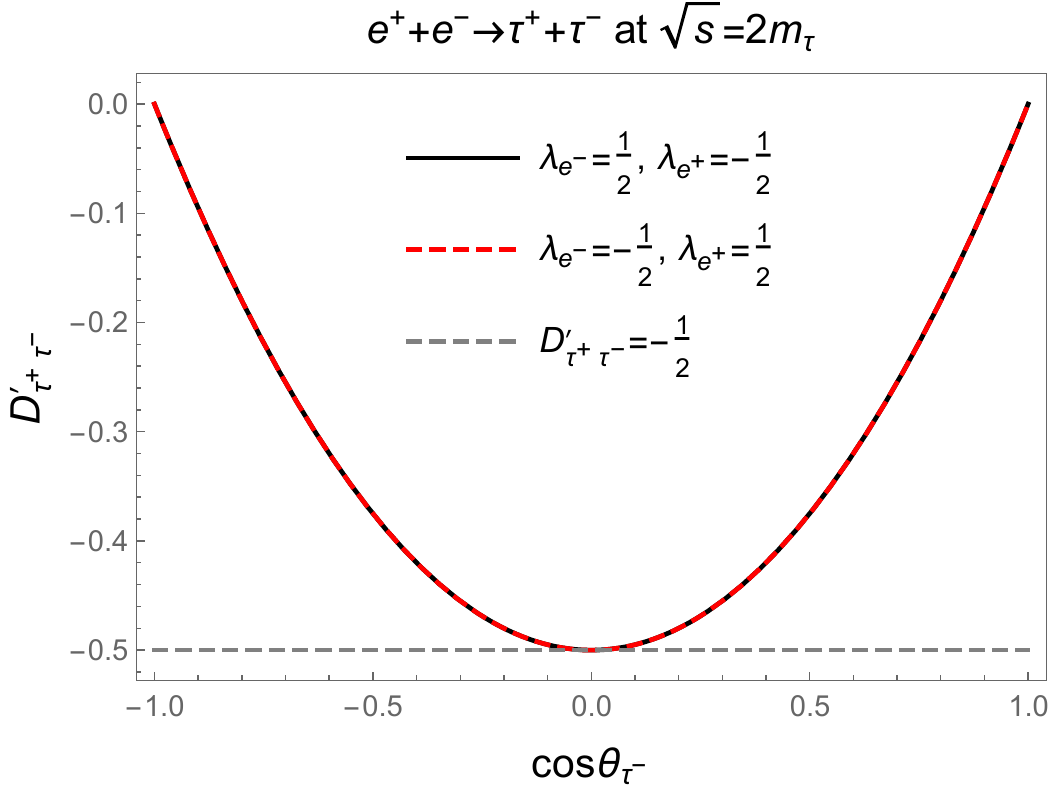} \\
\includegraphics[width=0.48\textwidth]{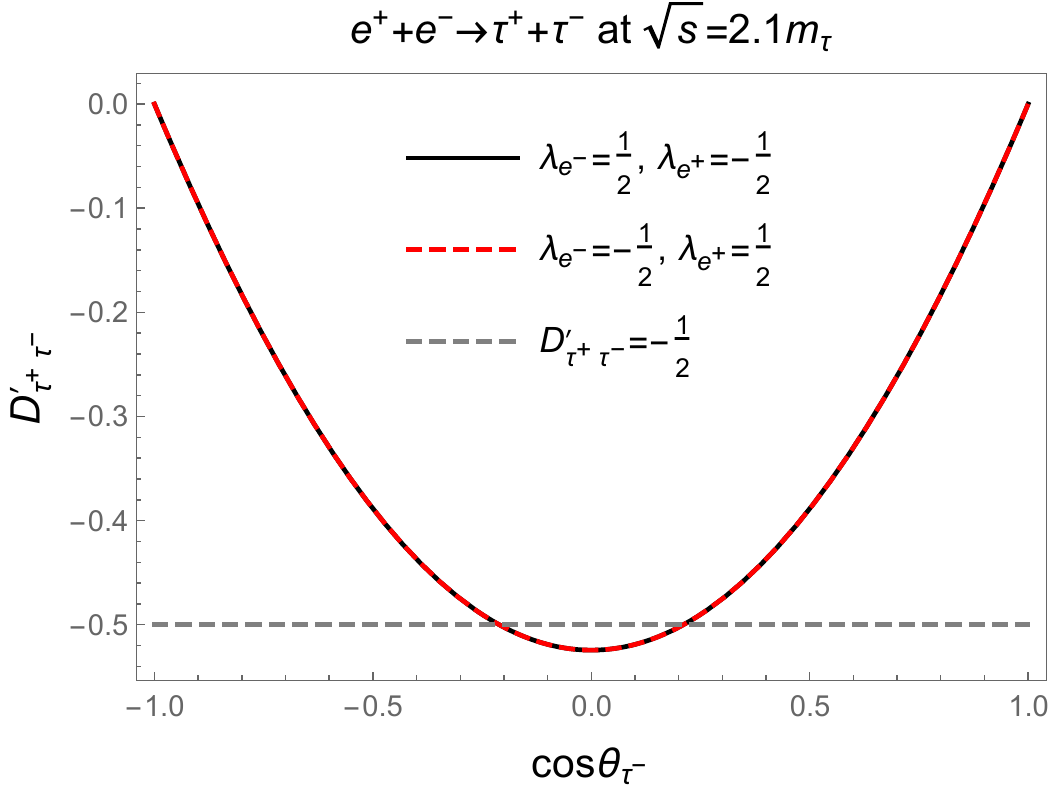}\hfill
\includegraphics[width=0.48\textwidth]{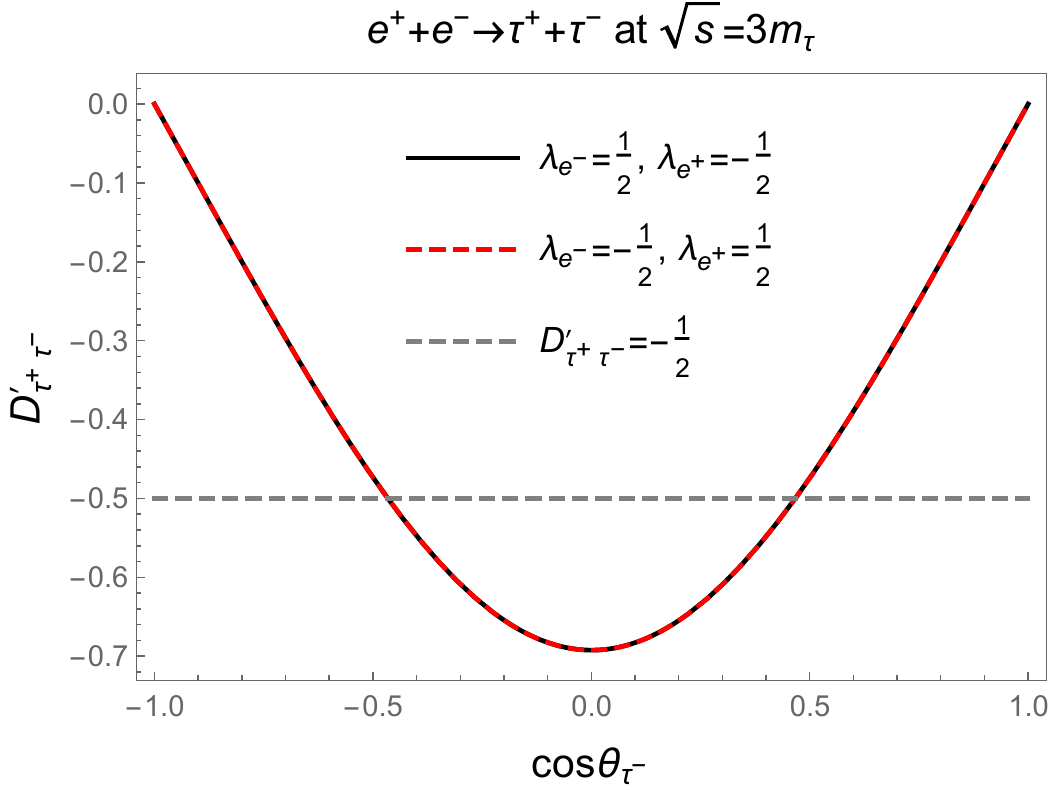}
\end{center}
\caption{The LO predictions of $ D^\prime_{\tau^+\tau^-} = \langle \cos\left(\phi_{e^+}-\phi_{e^-}\right)\rangle/C^\prime_{\tau^+ \tau^-} $ for $ \tau^+\tau^- $ pairs produced at $ e^+ e^- $ collider with $\sqrt{s}=2m_\tau$, $2.1m_\tau$, and $3m_\tau$, respectively. The angle $ \theta_{\tau^-} $ represents the polar angle of $ \tau^- $ in the laboratory frame. The symbols $ \lambda_{e^\pm} $ indicate the helicities of the $ e^+ $ and $ e^- $ beams, defined along their respective momentum directions in the laboratory frame.
\label{dppp}}
\end{figure}

Figure \ref{dppp} displays $ D^\prime_{\tau^+\tau^-} = \langle \cos\left(\phi_{e^+}-\phi_{e^-}\right)\rangle/C^\prime_{\tau^+ \tau^-} $ derived from $\tau^+\tau^-$ production at various $e^+ e^-$ collider energies, incorporating different initial-state polarization configurations. Our analysis reveals that above the $\tau^+\tau^-$ production threshold ($2m_\tau$), specific angular regions of $\tau^+\tau^-$ pair production (characterized by $\theta_{\tau^-}$ ranges) yield $D^\prime_{\tau^+\tau^-} < -\frac{1}{2}$, thus serving as conclusive evidence for quantum entanglement in $\tau^+\tau^-$ systems.

\section{Other examples}\label{sec16}

Next, we will investigate the characterization of quantum states in general multi-particle systems and develop criteria for assessing QE within these systems. We use the top quark ($t$), which has two polarization states, and the $ W $ boson, with its three polarization states, as examples to present the calculation results related to the QE of  general multi-particle systems. As for spin projection quantum numbers, we use $k=\pm \frac{1}{2}$ for fermions and $k=-1,0,1$ for massive vector particles. 

\subsection{$t\bar{t}$}

\begin{itemize}
\item For the processes $t \to W^+ + b$ and $\bar{t} \to W^- + \bar{b}$, let $\theta_{W^+ W^-}$ be the angle between the directions of the final-state $W^+$ and $W^-$ in the respective rest frames of $t$ and $\bar{t}$. Calculations at the LO level yield
\begin{align}
    &\langle \cos \theta_{W^+ W^-} \rangle \nonumber\\
    &= C_t^2\left( |\alpha_{\frac{1}{2},\frac{1}{2}}|^2 + |\alpha_{-\frac{1}{2},-\frac{1}{2}}|^2 - |\alpha_{\frac{1}{2},-\frac{1}{2}}|^2 - |\alpha_{-\frac{1}{2},\frac{1}{2}}|^2 + 2\alpha_{\frac{1}{2},\frac{1}{2}}\alpha^*_{-\frac{1}{2},-\frac{1}{2}} + 2\alpha_{-\frac{1}{2},-\frac{1}{2}}\alpha^*_{\frac{1}{2},\frac{1}{2}} \right)\label{jjjggg}\\
    &\in C_t^2 \times [-1, 3]~,
\end{align}
where
\begin{align}
   & C_t=\frac{1}{3}\frac{q \left({m}^2_t-{m}^2_b-2 {m}^2_W\right)}{3 {m}^2_W \sqrt{{m}^2_b+q^2}+2 {m}_t q^2}~,\\
   & q=\frac{m_t}{2}\sqrt{\left(1-\left(\frac{{m_W}+{m_b}}{{m_t}}\right)^2\right) \left(1-\left(\frac{{m_W}-{m_b}}{{m_t}}\right)^2\right)}~.
\end{align}
When Eq.~(\ref{a2}) is satisfied, we have
\begin{align}
    \langle \cos \theta_{W^+ W^-} \rangle \in C_t^2 \times [-1, 1]~.
\end{align}
Therefore, for these processes, the sufficient condition for QE between $t$ and $\bar{t}$ is
\begin{align}
    \langle \cos \theta_{W^+ W^-} \rangle \in C_t^2 \times (1, 3]~.
\end{align}

\item For the processes $t \to W^+ + b$ and $\bar{t} \to e^- + \bar{\nu}_e + \bar{b}$, let $\theta_{W^+ e^-}$ be the angle between the directions of the final-state $W^+$ and $e^-$ in the respective rest frames of $t$ and $\bar{t}$. Calculations at the LO level give
\begin{align}
    &\langle \cos \theta_{W^+ e^-} \rangle \nonumber\\
    &= \frac{C_t}{3} \times \left( |\alpha_{\frac{1}{2},\frac{1}{2}}|^2 + |\alpha_{-\frac{1}{2},-\frac{1}{2}}|^2 - |\alpha_{\frac{1}{2},-\frac{1}{2}}|^2 - |\alpha_{-\frac{1}{2},\frac{1}{2}}|^2 + 2\alpha_{\frac{1}{2},\frac{1}{2}}\alpha^*_{-\frac{1}{2},-\frac{1}{2}} + 2\alpha_{-\frac{1}{2},-\frac{1}{2}}\alpha^*_{\frac{1}{2},\frac{1}{2}} \right)\label{jjjgggg}\\
    &\in \frac{C_t}{3} \times [-1, 3]~.
\end{align}
When Eq.~(\ref{a2}) is satisfied, we find
\begin{align}
    \langle \cos \theta_{W^+ e^-} \rangle \in \frac{C_t}{3} \times [-1, 1]~.
\end{align}
Thus, in this case, the sufficient condition for QE between $t$ and $\bar{t}$ is
\begin{align}
    \langle \cos \theta_{W^+ e^-} \rangle \in \frac{C_t}{3} \times (1, 3]~.
\end{align}
\end{itemize}

It is noteworthy that the results in Eqs.~(\ref{jjgg}), (\ref{jjjggg}), and (\ref{jjjgggg}) share the same structural form in their expressions, differing only in the magnitudes of their leading constant factors.

\subsection{$W^- W^+$}

The complete polarization state of a $W^-W^+$ system can be expressed through the quantum superposition:
\begin{align}
&|W^-W^+\rangle=\sum_{k,j=-1,0,1}\alpha_{k,j}|k\rangle_{W^-}|j\rangle_{W^+}~,\\
&\sum_{k,j=-1,0,1}\left|\alpha_{k,j}\right|^2=1~.
\end{align}

\begin{itemize}
    \item For the processes $W^- \to e^- + \bar{\nu}_e$ and $W^+ \to e^+ + \nu_e$, let us consider $\theta_{e^- e^+}$ as the angle between the directions of the final-state $e^-$ and $e^+$ in the rest frames of $W^-$ and $W^+$, respectively. Calculations at the LO level yield
\begin{align}
    \langle \cos \theta_{e^- e^+} \rangle =& \frac{1}{4} \times \left( |\alpha_{-1,-1}|^2 + |\alpha_{1,1}|^2 - |\alpha_{-1,1}|^2 - |\alpha_{1,-1}|^2 - \alpha_{-1,-1}\alpha^*_{0,0} - \alpha_{0,0}\alpha^*_{-1,-1} \right. \nonumber\\
    &\left. - \alpha_{1,1}\alpha^*_{0,0} - \alpha_{0,0}\alpha^*_{1,1} - \alpha_{-1,0}\alpha^*_{0,1} - \alpha_{0,1}\alpha^*_{-1,0} - \alpha_{0,-1}\alpha^*_{1,0} - \alpha_{1,0}\alpha^*_{0,-1} \right) \label{ww123}\\
    \in & \, [-\frac{1}{4}, \frac{1}{2}]~.
\end{align}
When the condition analogous to Eq.~(\ref{a2}) is satisfied, we have
\begin{align}
    \langle \cos \theta_{e^- e^+} \rangle \in [-\frac{1}{4}, \frac{1}{4}]~.
\end{align}
Thus, in this context, the sufficient criterion for QE between $W^-$ and $W^+$ is
\begin{align}
    \langle \cos \theta_{e^- e^+} \rangle \in (\frac{1}{4}, \frac{1}{2}]~.
\end{align}

\begin{figure}[tbp]
\begin{center}
\includegraphics[width=0.48\textwidth]{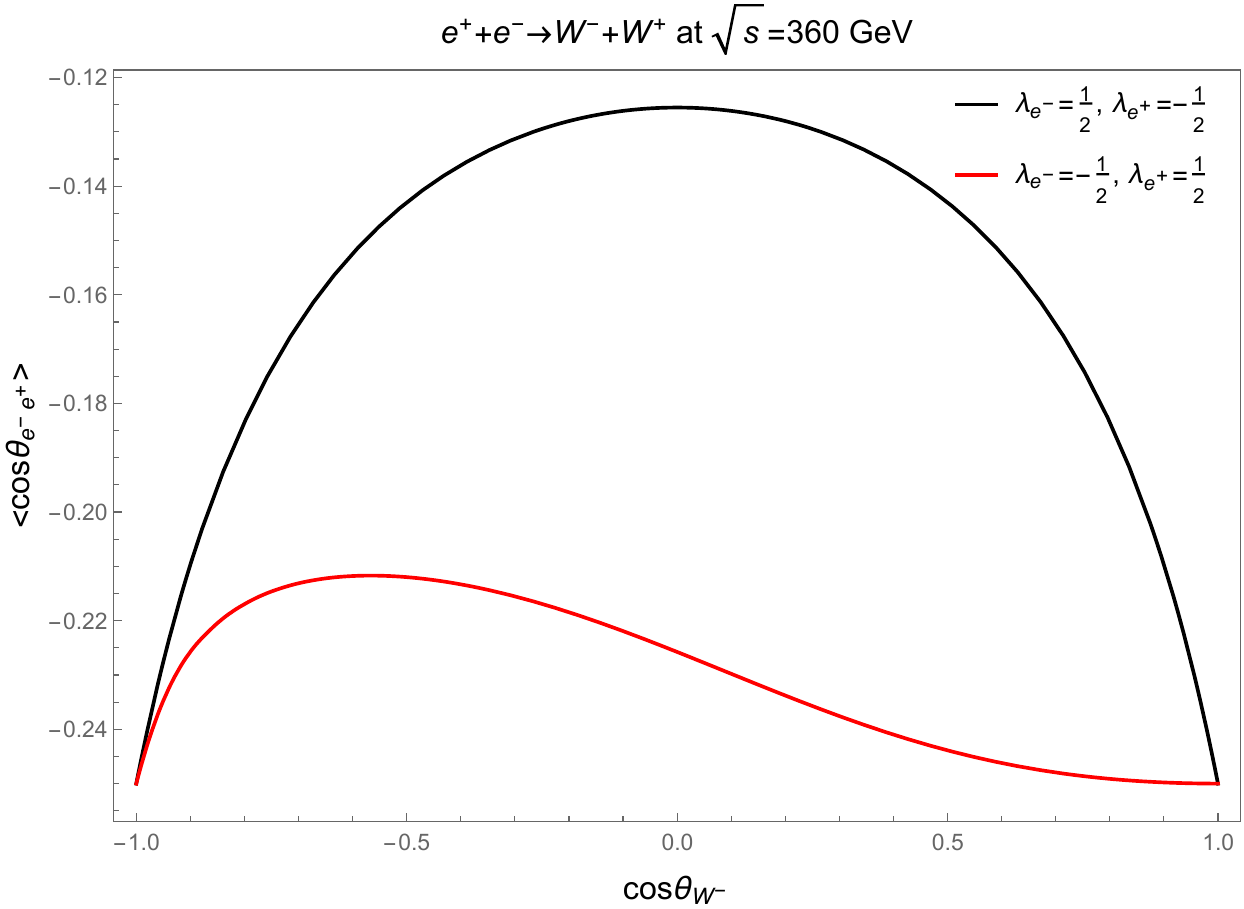} 
\includegraphics[width=0.48\textwidth]{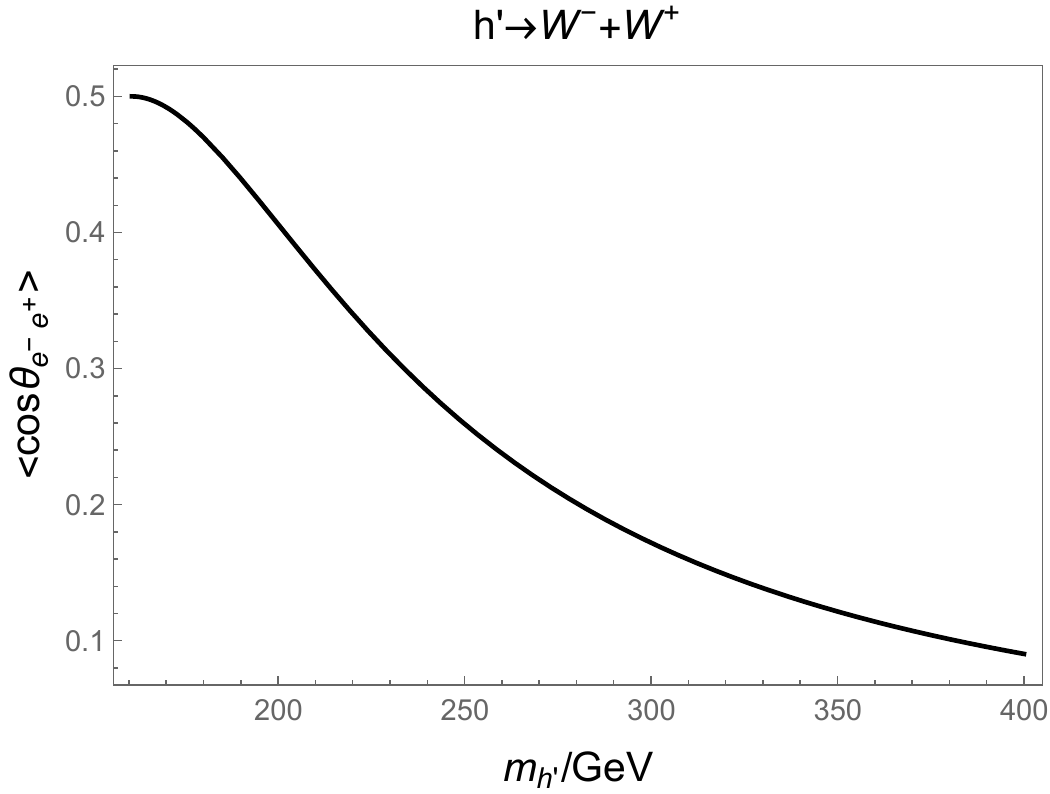} 
\end{center}
\caption{Left: The LO predictions of $\langle \cos \theta_{e^- e^+} \rangle$ for the $W^- W^+$ pair produced at $ e^+ e^- $ collider with $\sqrt{s}=360$ GeV. The angle $ \theta_{W^-} $ represents the polar angle of $ W^- $ in the laboratory frame. The symbols $ \lambda_{e^\pm} $ indicate the helicities of the $ e^+ $ and $ e^- $ beams, defined along their respective momentum directions in the laboratory frame.
Right: The LO predictions of $\langle \cos \theta_{e^- e^+} \rangle$ for the $W^- W^+$ pair produced from $h^\prime$ decay.
\label{wwpm}}
\end{figure}

Without considering any BSM particles or interactions, the left panel of Figure \ref{wwpm} presents the LO predictions of $\langle \cos \theta_{e^- e^+} \rangle$ as described in Eq.~(\ref{ww123}) for $ W^- W^+ $ pairs produced at an $ e^+ e^- $ collider with a c.m. energy of $\sqrt{s}=360$ GeV. Different polarization states of the $ e^+ $ and $ e^- $ beams are taken into account. Across different polar angles of the $ W^- $, the value of $\langle \cos \theta_{e^- e^+} \rangle$ consistently falls below zero. This trend holds true for other beam energies as well, indicating insufficient evidence for QE in the $ W^- W^+ $ system.

\item We consider a BSM Higgs-like particle, $ h^\prime $, with a mass greater than twice that of the $ W $ boson. The interaction between $ h^\prime $ and the $ W $ bosons is analogous to the interaction form of the SM Higgs boson, represented as
\begin{align}
    \propto h^\prime g^{\mu\nu} W^-_{\mu} W^+_{\nu}~.
\end{align}
For the processes $W^- \to e^- + \bar{\nu}_e$ and $W^+ \to e^+ + \nu_e$, the right panel of Figure \ref{wwpm} showcases the LO predictions of $\langle \cos \theta_{e^- e^+} \rangle$ in Eq.~(\ref{ww123}) for $ W^- W^+ $ pairs arising from the decay of $ h^\prime $. As the mass of $ h^\prime $, denoted $ m_{h^\prime} $, approaches the threshold for $ W^- W^+ $ pair production, which is twice the mass of the $ W $ boson, the value of $\langle \cos \theta_{e^- e^+} \rangle$ increases. Specifically, $\langle \cos \theta_{e^- e^+} \rangle$ exceeds $\frac{1}{4}$ when $ m_{h^\prime} < 254.1 $ GeV, which can provide sufficient evidence for QE in the $ W^- W^+ $ system. Therefore, measuring $\langle \cos \theta_{e^- e^+} \rangle$ in Eq.~(\ref{ww123}) for $ W^- W^+ $ pairs offers a model-independent approach to searching for $ h^\prime $ with $ m_{h^\prime} < 254.1 $ GeV. We intend to pursue this search, along with the accompanying background analysis, in our future work.

\item For the processes $W^- \to e^- + \bar{\nu}_e$ and $W^+ \to e^+ + \nu_e$, the calculation at the LO level yields  
\begin{align}
    \langle \cos\left(2\phi_{e^+}-2\phi_{e^-}\right)\rangle = \frac{1}{8}\left(\alpha_{-1,1}\alpha_{1,-1}^* + \alpha_{1,-1}\alpha_{-1,1}^*\right),
\end{align}  
where $\phi_{e^-}$ and $\phi_{e^+}$ denote the azimuthal angles of $e^-$ and $e^+$ in the rest frames of $W^-$ and $W^+$, respectively, defined following the geometric convention established in Section~{\ref{211}}.
So, we define the normalized entanglement witness  
\begin{align}
    D^\prime_{W^-W^+} = 8\langle \cos\left(2\phi_{e^+}-2\phi_{e^-}\right)\rangle = \alpha_{-1,1}\alpha_{1,-1}^* + \alpha_{1,-1}\alpha_{-1,1}^*~.
\end{align}  
The existence of QE in $W^+ W^-$ pairs is conclusively demonstrated when  
\begin{align}
   D^\prime_{W^-W^+} \in [-1, -\tfrac{1}{2}) \cup (\tfrac{1}{2}, 1]~.
\end{align}

\begin{figure}[tbp]
\begin{center}
\includegraphics[width=0.48\textwidth]{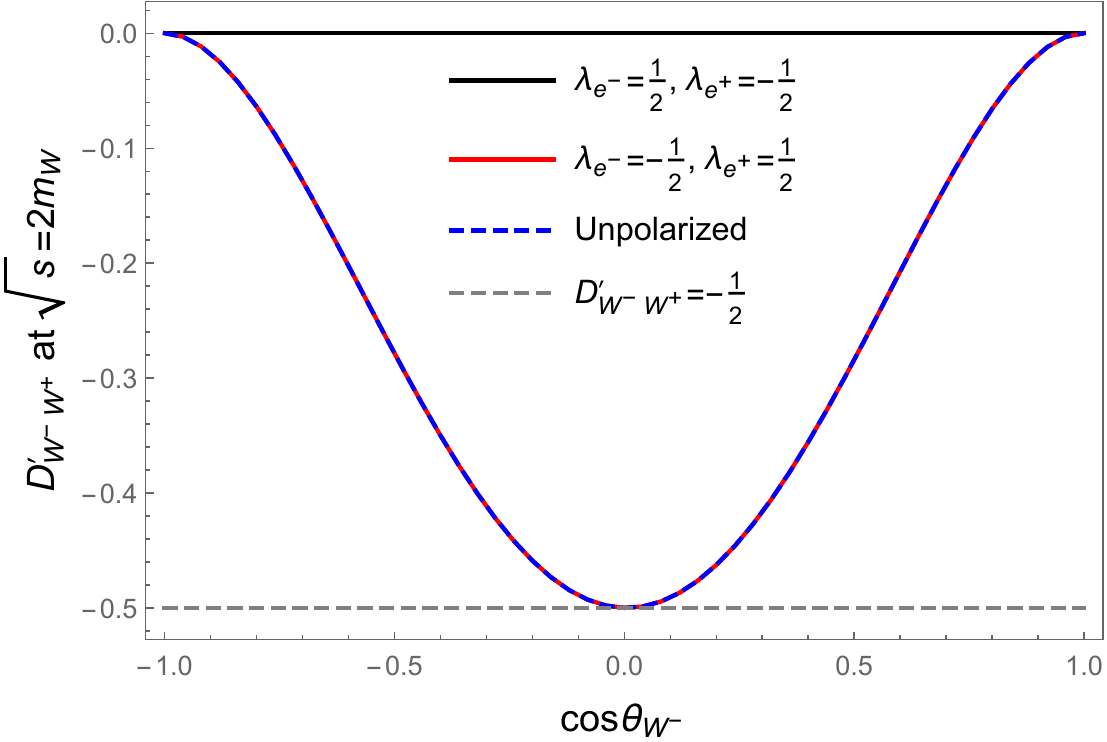} \\
\includegraphics[width=0.48\textwidth]{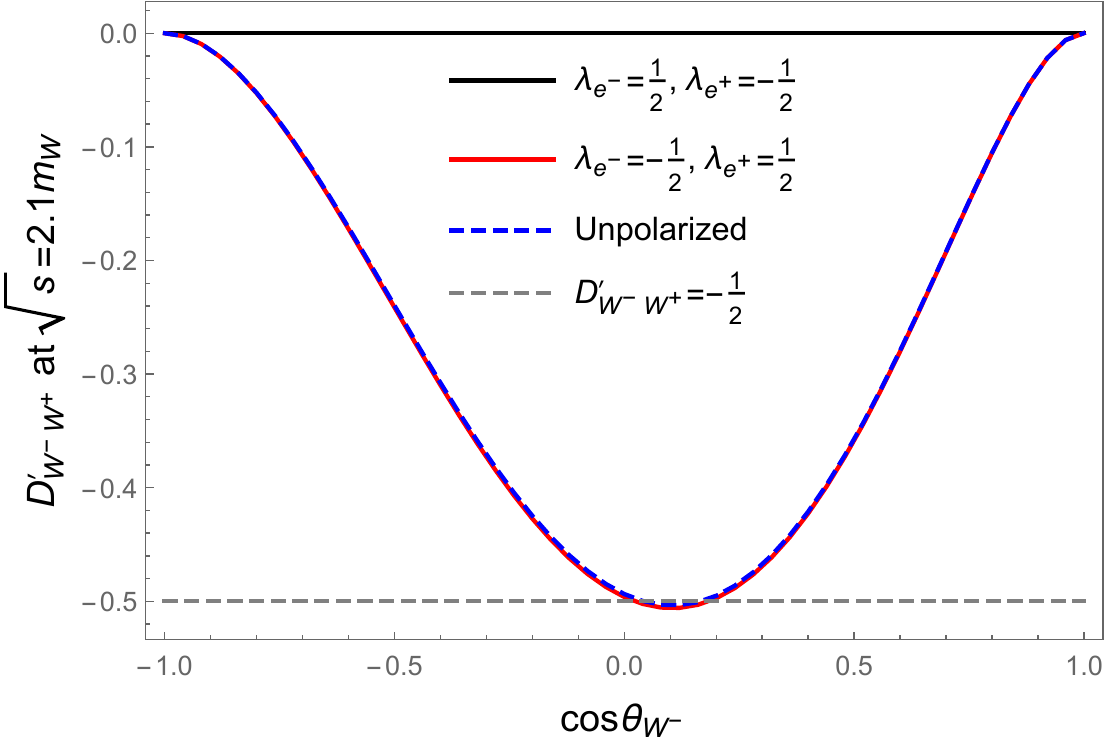}\hfill
\includegraphics[width=0.48\textwidth]{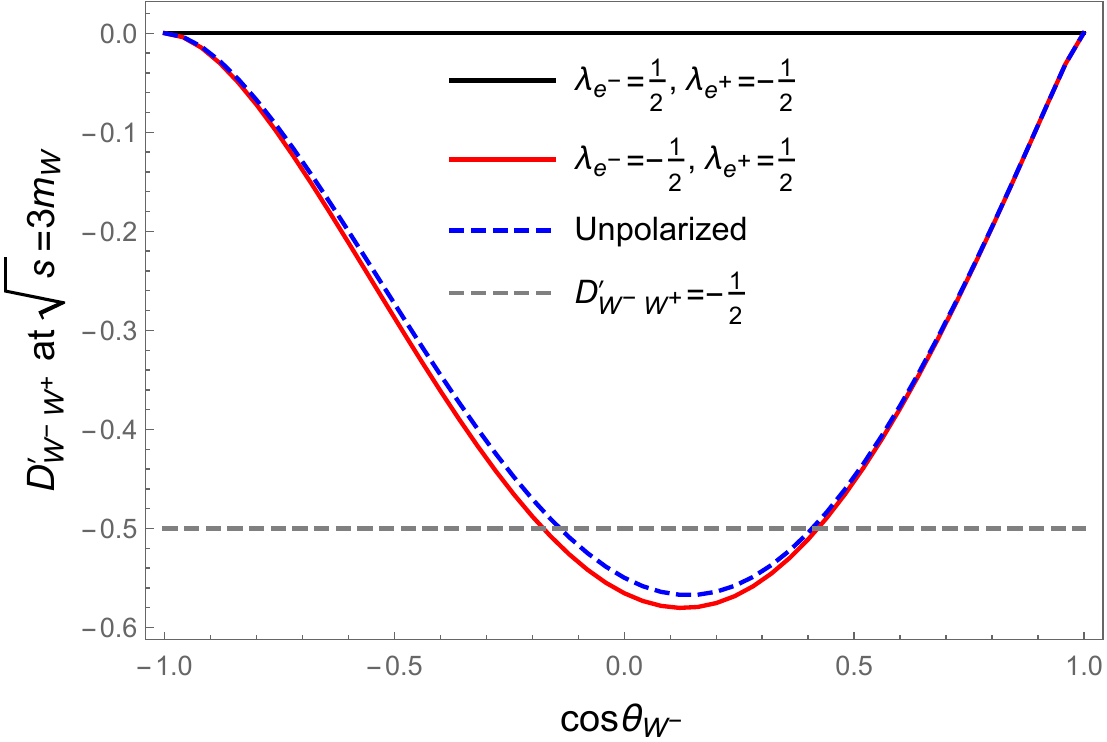}
\end{center}
\caption{The LO predictions of $D^\prime_{W^-W^+} = 8\langle \cos\left(2\phi_{e^+}-2\phi_{e^-}\right)\rangle$ for $ W^+ W^- $ pairs produced at $ e^+ e^- $ collider with $\sqrt{s}=2m_W$, $2.1m_W$ GeV, and $3m_W$, respectively. The angle $ \theta_{W^-} $ represents the polar angle of the top quark $ W^- $ in the laboratory frame. The symbols $ \lambda_{e^\pm} $ indicate the helicities of the $ e^+ $ and $ e^- $ beams, defined along their respective momentum directions in the laboratory frame.
\label{ddppp}}
\end{figure}
\end{itemize}

Figure \ref{ddppp} displays the $D^\prime_{W^-W^+} = 8\langle \cos\left(2\phi_{e^+}-2\phi_{e^-}\right)\rangle$ derived from $W^+ W^-$ production at various $e^+ e^-$ collider energies, incorporating different initial-state polarization configurations. Our analysis reveals that above the $W^+ W^-$ production threshold ($2m_W$), specific angular regions of $W^+ W^-$ pair production (characterized by $\theta_{W^-}$ ranges) yield $D^\prime_{W^-W^+} < -\frac{1}{2}$, thus serving as conclusive evidence for quantum entanglement in $W^+ W^-$ systems.

\subsection{$W^- t$}

The complete polarization state of a $W^- t$ system can be expressed through the quantum superposition:
\begin{align}
&|W^- t\rangle=\sum_{k=-1,0,1; ~j=\pm\frac{1}{2}
}\alpha_{k,j}|k\rangle_{W^-}|j\rangle_{t}~,\\
&\sum_{k=-1,0,1; ~j=\pm\frac{1}{2}
}\left|\alpha_{k,j}\right|^2=1~.\\
\end{align}
For processes of $W^- \to e^- + \bar{\nu}_e$ and $t \to W^+ + b$ and $\theta_{e^- W^+}$ being the angle between the directions of the final-state $e^-$ and $W^+$ in the rest frames of $W^-$ and $t$, respectively, calculations at the LO level give
\begin{align}
    \langle \cos \theta_{e^- W^+} \rangle =& \frac{C_t}{2}\times\left( |\alpha_{-1,-\frac{1}{2}}|^2+|\alpha_{1,\frac{1}{2}}|^2-|\alpha_{-1,\frac{1}{2}}|^2-|\alpha_{1,-\frac{1}{2}}|^2\right. \nonumber\\
    &\left. +\sqrt{2}\left(\alpha_{-1,-\frac{1}{2}}\alpha^*_{0,\frac{1}{2}}+\alpha_{0,\frac{1}{2}}\alpha^*_{-1,-\frac{1}{2}}+\alpha_{0,-\frac{1}{2}}\alpha^*_{1,\frac{1}{2}}+\alpha_{1,\frac{1}{2}}\alpha^*_{0,-\frac{1}{2}}\right)\right)\label{pi22} \\
     \in & \frac{C_t}{2}\times [-1,2]~.\label{422}
\end{align}
When the condition parallel to Eq.~(\ref{a2}) is satisfied, we get
\begin{align}
    \langle \cos \theta_{e^- W^+} \rangle  \in \frac{C_t}{2}\times[-1,1]~.
\end{align}
So, in this case the sufficient criterion of QE between $W^-$ and $\bar{t}$ is
\begin{align}
   D^\prime_{W^- t} =\frac{2}{C_t}\langle \cos \theta_{e^- W^+} \rangle  \in  (1,2]~.
\end{align}

\begin{figure}[tbp]
\begin{center}
\includegraphics[width=0.48\textwidth]{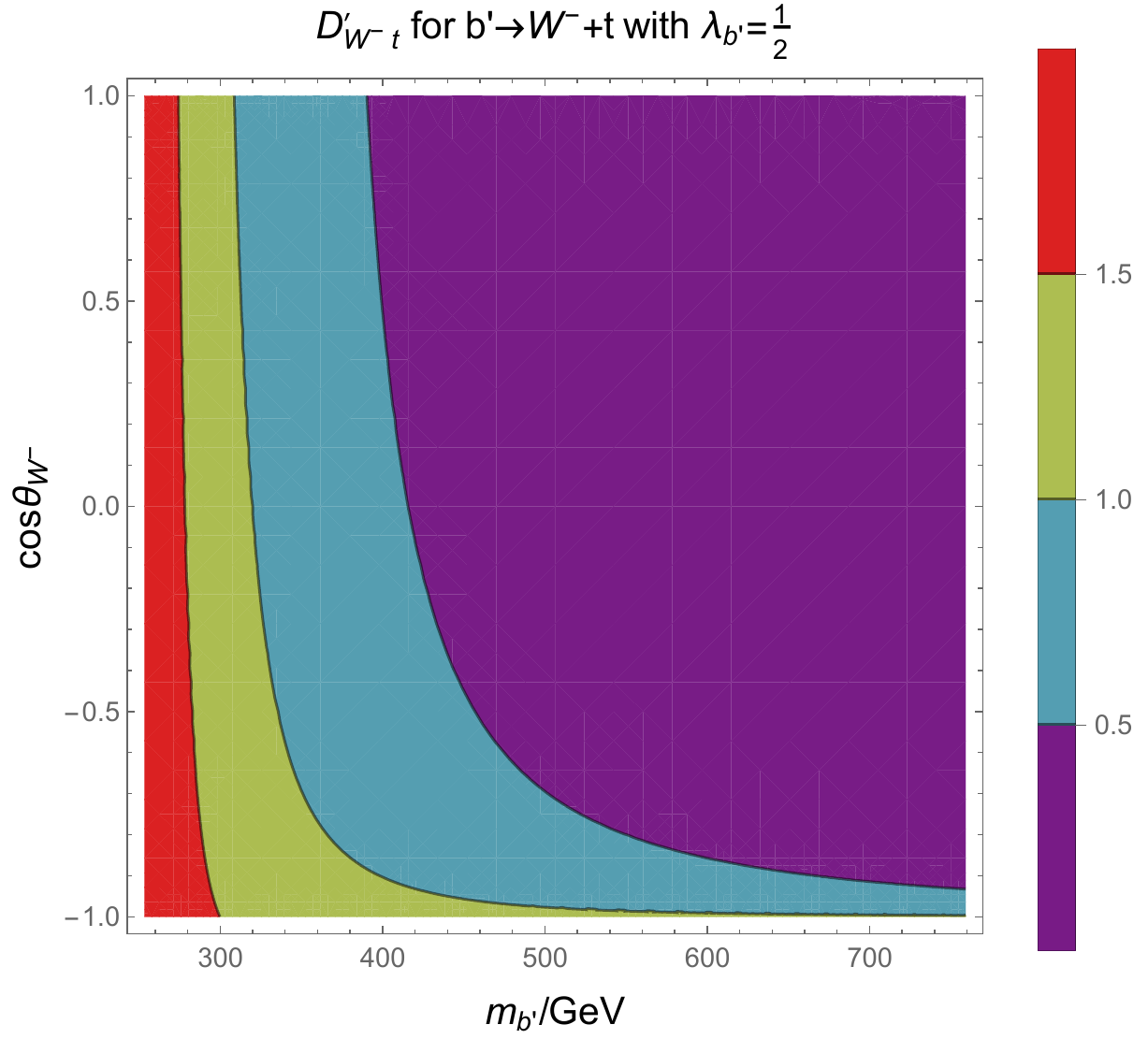} 
\includegraphics[width=0.48\textwidth]{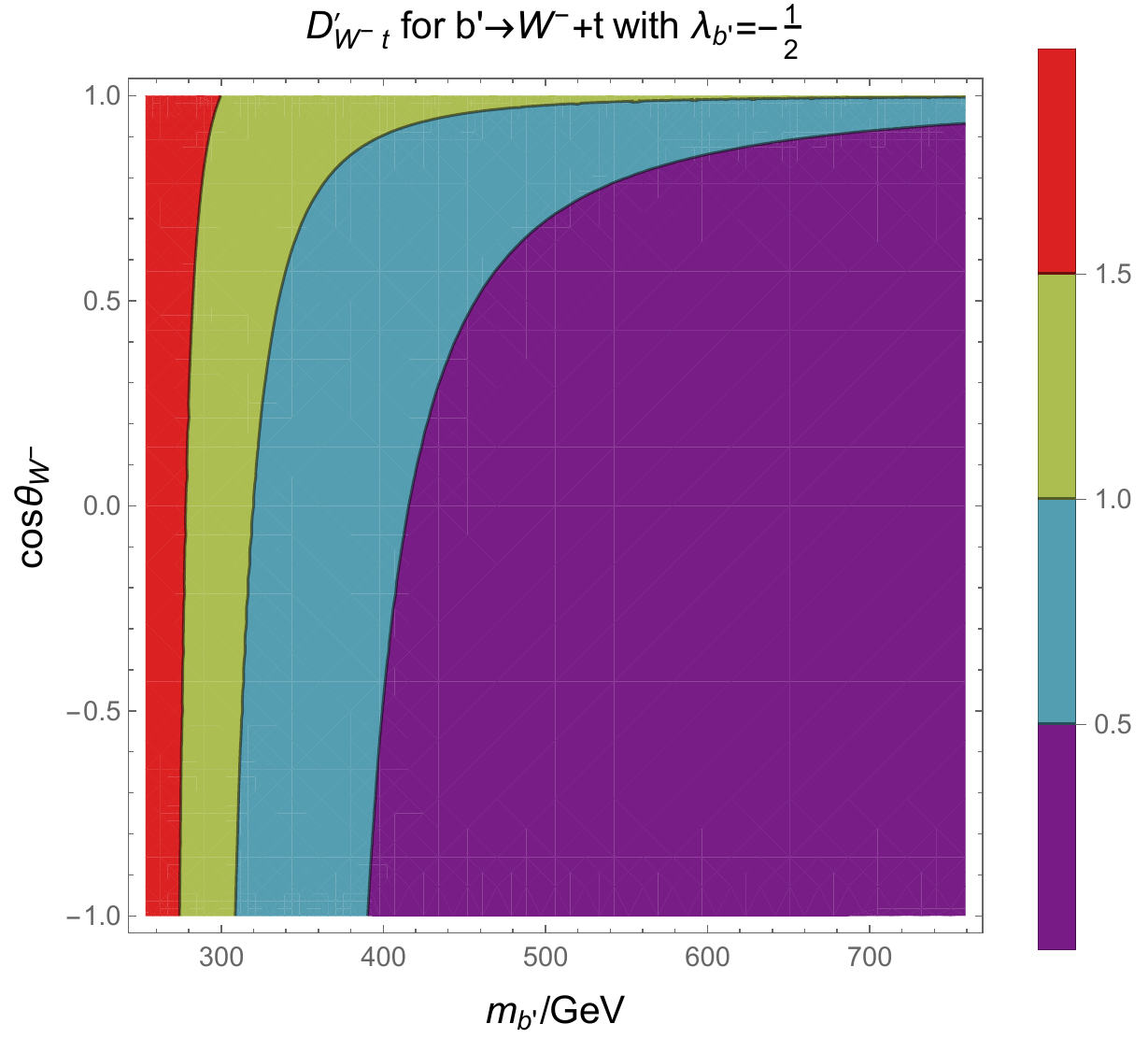}\\
\includegraphics[width=0.48\textwidth]{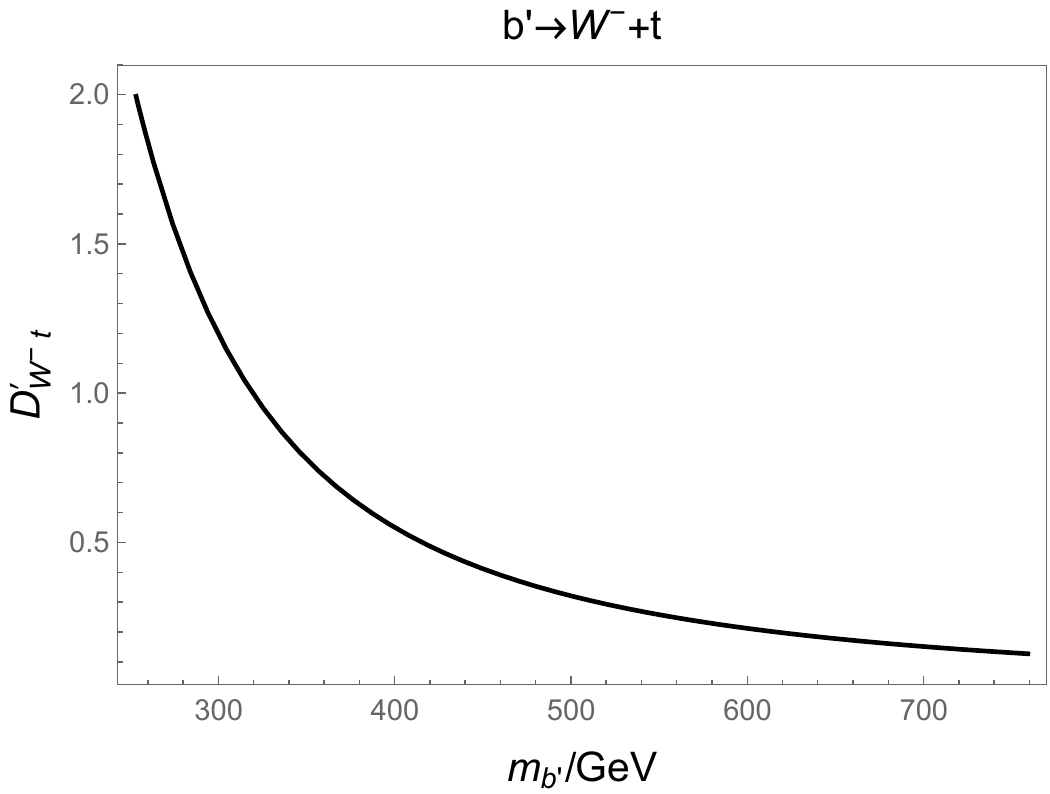}
\end{center}
\caption{Upper: The LO predictions of $D^\prime_{W^- t} =\frac{2}{C_t}\langle \cos \theta_{e^- W^+} \rangle$ for the $W^- t$ pair produced by $b^\prime$ decay. Here, $\theta_{W^-}$ represents the angle between the momentum direction of the $W^-$ in the $b^\prime$ rest frame and the direction of $b^\prime$ motion in the laboratory frame. The symbols $\lambda_{b^\prime}$ denote the helicities of $b^\prime$, defined along the momentum direction of $b^\prime$ in the laboratory frame.
Lower: The LO predictions of $D^\prime_{W^- t} =\frac{2}{C_t}\langle \cos \theta_{e^- W^+} \rangle$ for the $W^- t$ produced by $b^\prime$ decay, averaged over $\cos\theta_{W^-}$.
\label{bpp}}
\end{figure}

We investigate a BSM bottom-like quark, denoted as $ b^\prime $, which has a mass heavier than $ m_W + m_t $. The interaction of $ b^\prime $ with the $ W $ boson and the top quark $ t $ is analogous to that of the SM bottom quark $ b $. This interaction can be expressed as follows
\begin{align}
    \propto W^+_\mu \bar{t}_\text{L} \gamma^\mu b^\prime_\text{L} + h.c.
\end{align}
The upper panels of Figure \ref{bpp} present the LO predictions of $D^\prime_{W^- t} =\frac{2}{C_t}\langle \cos \theta_{e^- W^+} \rangle$ as derived from Eq.~(\ref{422}) for the $ W^- t $ pair produced by the decay of $ b^\prime $. As the mass of $ b^\prime $, denoted $ m_{b^\prime} $, approaches the threshold of $ m_W + m_t $, we observe an increase in the value of $D^\prime_{W^- t}$. Furthermore, it is evident that $D^\prime_{W^- t}$ is dependent on $ \theta_{W^-} $, which represents the angle between the momentum direction of the $ W^- $ in the rest frame of $ b^\prime $ and the direction of $ b^\prime $ motion in the laboratory frame. 
After averaging over $ \cos\theta_{W^-} $, the LO predictions of $D^\prime_{W^- t}$ are displayed in the lower panel of Figure \ref{bpp}. Notably, $D^\prime_{W^- t}$ exceeds $ 1 $ when $ m_{b^\prime} < 319.5 $ GeV, providing compelling evidence for QE in the $ W^- t $ system. Therefore, measuring $D^\prime_{W^- t}$ as indicated in Eq.~(\ref{422}) for the $ W^- t $ pairs presents a model-independent strategy for searching for $ b^\prime $ with $ m_{b^\prime} < 319.5 $ GeV. We plan to pursue this search, along with a detailed background analysis, in our future work.

\subsection{$ttt$}

We characterize the polarization state of the $ttt$ system through the quantum superposition:
\begin{align}
    &\left|ttt\right\rangle=\sum_{k,j,m=\pm \frac{1}{2}}\alpha_{k,j,m}|k\rangle_{t_1}|j\rangle_{t_2}|m\rangle_{t_3}~,\\
    &\sum_{k,j,m=\pm \frac{1}{2}}\left|\alpha_{k,j,m}\right|^2=1~.
\end{align}

Examining the decay process $t_i\to W^+_i +b_i$ for each top quark ($i=1,2,3$), we define the spherical coordinates $(\theta_i,\phi_i)$ for the $W^+_i$ momentum direction in respective $t_i$ rest frames following Section~\ref{multi}. The momentum unit vectors are expressed as
\begin{align}
    \hat{e}_{W^+_i}=(\sin\theta_i\cos\phi_i,\sin\theta_i\sin\phi_i,\cos\theta_i).
\end{align}
The triple product correlation observable is constructed as
\begin{align}
    &\left(\hat{e}_{W^+_1}\times\hat{e}_{W^+_2}\right)\cdot \hat{e}_{W^+_3}\nonumber\\
    &= -\sin\theta_1\sin\theta_2\cos\theta_3\sin(\phi_1-\phi_2)-\sin\theta_2\sin\theta_3\cos\theta_1\sin(\phi_2-\phi_3)-\sin\theta_3\sin\theta_1\cos\theta_2\sin(\phi_3-\phi_1)~.
\end{align}
The explicit computation at the LO level reveals:
\begin{align}
&\left\langle\left(\hat{e}_{W^+_1}\times\hat{e}_{W^+_2}\right)\cdot \hat{e}_{W^+_3}\right\rangle \nonumber\\
    =& 2C_t^{3}~i\times\left(-\alpha_{-\frac{1}{2},-\frac{1}{2},\frac{1}{2}}\alpha_{-\frac{1}{2},\frac{1}{2},-\frac{1}{2}}^* +\alpha^*_{-\frac{1}{2},-\frac{1}{2},\frac{1}{2}}\alpha_{-\frac{1}{2},\frac{1}{2},-\frac{1}{2}}
    +\alpha_{-\frac{1}{2},-\frac{1}{2},\frac{1}{2}}\alpha_{\frac{1}{2},-\frac{1}{2},-\frac{1}{2}}^* -\alpha^*_{-\frac{1}{2},-\frac{1}{2},\frac{1}{2}}\alpha_{\frac{1}{2},-\frac{1}{2},-\frac{1}{2}} \right. \nonumber\\
   &  -\alpha_{-\frac{1}{2},\frac{1}{2},-\frac{1}{2}}\alpha_{\frac{1}{2},-\frac{1}{2},-\frac{1}{2}}^* +\alpha^*_{-\frac{1}{2},\frac{1}{2},-\frac{1}{2}}\alpha_{\frac{1}{2},-\frac{1}{2},-\frac{1}{2}}
     +\alpha_{-\frac{1}{2},\frac{1}{2},\frac{1}{2}}\alpha_{\frac{1}{2},-\frac{1}{2},\frac{1}{2}}^* -\alpha^*_{-\frac{1}{2},\frac{1}{2},\frac{1}{2}}\alpha_{\frac{1}{2},-\frac{1}{2},\frac{1}{2}} \nonumber\\
     & \left. -\alpha_{-\frac{1}{2},\frac{1}{2},\frac{1}{2}}\alpha_{\frac{1}{2},\frac{1}{2},-\frac{1}{2}}^* + \alpha^*_{-\frac{1}{2},\frac{1}{2},\frac{1}{2}}\alpha_{\frac{1}{2},\frac{1}{2},-\frac{1}{2}}
     +\alpha_{\frac{1}{2},-\frac{1}{2},\frac{1}{2}}\alpha_{\frac{1}{2},\frac{1}{2},-\frac{1}{2}}^*- \alpha^*_{\frac{1}{2},-\frac{1}{2},\frac{1}{2}}\alpha_{\frac{1}{2},\frac{1}{2},-\frac{1}{2}} \right) \nonumber\\
     \in & ~2C_t^{3}\times[-\sqrt{3},\sqrt{3}]
\end{align}
When the following separability condition holds:
\begin{align}
    \alpha_{k,j,m}=\beta_k \gamma_j \kappa_m~,\quad k,j,m=\pm\frac{1}{2}~,
\end{align}
the system exhibits no quantum entanglement, yielding:
\begin{align}
&\left\langle\left(\hat{e}_{W^+_1}\times\hat{e}_{W^+_2}\right)\cdot \hat{e}_{W^+_3}\right\rangle\in 2C_t^{3}\times \left[-\frac{1}{2},\frac{1}{2}\right]~.
\end{align}
Therefore, the sufficient criterion for genuine tripartite entanglement in the $ttt$ system is
\begin{align}
    D^\prime_{3t}=& \frac{1}{2C_t^{3}} \left\langle\left(\hat{e}_{W^+_1}\times\hat{e}_{W^+_2}\right)\cdot \hat{e}_{W^+_3}\right\rangle \in\left[-\sqrt{3},-\frac{1}{2}\right)\cup \left(\frac{1}{2},\sqrt{3}\right].
\end{align}

\subsection{$t\bar{t}W^-$}

We characterize the polarization state of the $t\bar{t}W^-$ system through the quantum superposition:
\begin{align}
    &|t\bar{t}W^-\rangle=\sum_{k,j=\pm \frac{1}{2};~m=-1,0,1} \alpha_{k,j,m}|k\rangle_{t} |j\rangle_{\bar{t}} |m\rangle_{W^-}~,\\
    &\sum_{k,j=\pm \frac{1}{2};~m=-1,0,1} \left|\alpha_{k,j,m}\right|^2=1~.
\end{align}

Considering the decay channels $t\to W^+ b$, $\bar{t}\to W^- \bar{b}$, and $W^- \to e^- \bar{\nu}_e$, we analyze the angular correlations of final-state particles ($W^+$, $W^-$, and $e^-$) in their respective parent particle rest frames. Following the methodology in Section~\ref{multi}, we define the momentum direction vectors:
\begin{align}
    \hat{e}_{W^+}=(\sin\theta_1\cos\phi_1,\sin\theta_1\sin\phi_1,\cos\theta_1)~, \\
    \hat{e}_{W^-}=(\sin\theta_2\cos\phi_2,\sin\theta_2\sin\phi_2,\cos\theta_2)~, \\
    \hat{e}_{e^-}=(\sin\theta_3\cos\phi_3,\sin\theta_3\sin\phi_3,\cos\theta_3)~,
\end{align}
and construct the triple product correlation observable:
\begin{align}
    &\left(\hat{e}_{W^+}\times\hat{e}_{W^-}\right)\cdot \hat{e}_{e^-}\nonumber\\
    &= -\sin\theta_1\sin\theta_2\cos\theta_3\sin(\phi_1-\phi_2)-\sin\theta_2\sin\theta_3\cos\theta_1\sin(\phi_2-\phi_3)-\sin\theta_3\sin\theta_1\cos\theta_2\sin(\phi_3-\phi_1)~.
\end{align}
The quantum expectation value calculation at the LO level reveals:
\begin{align}
&\left\langle\left(\hat{e}_{W^+}\times\hat{e}_{W^-}\right)\cdot \hat{e}_{e^-}\right\rangle \nonumber\\
    =& \frac{C_t^{2}}{\sqrt{2}}~i\times\left(
    -\alpha_{-\frac{1}{2},-\frac{1}{2},0}\alpha_{-\frac{1}{2},\frac{1}{2},-1}^*
    +\alpha^*_{-\frac{1}{2},-\frac{1}{2},0}\alpha_{-\frac{1}{2},\frac{1}{2},-1}
    -\alpha_{-\frac{1}{2},-\frac{1}{2},1}\alpha_{-\frac{1}{2},\frac{1}{2},0}^*
    +\alpha^*_{-\frac{1}{2},-\frac{1}{2},1}\alpha_{-\frac{1}{2},\frac{1}{2},0}
    \right. \nonumber\\
& +\alpha_{-\frac{1}{2},-\frac{1}{2},0}\alpha_{\frac{1}{2},-\frac{1}{2},-1}^*
  -\alpha^*_{-\frac{1}{2},-\frac{1}{2},0}\alpha_{\frac{1}{2},-\frac{1}{2},-1}
  -\sqrt{2}\alpha_{-\frac{1}{2},\frac{1}{2},-1}\alpha_{\frac{1}{2},-\frac{1}{2},-1}^*
+\sqrt{2}\alpha^*_{-\frac{1}{2},\frac{1}{2},-1}\alpha_{\frac{1}{2},-\frac{1}{2},-1}\nonumber\\
& +\alpha_{-\frac{1}{2},-\frac{1}{2},1}\alpha_{\frac{1}{2},-\frac{1}{2},0}^*
-\alpha^*_{-\frac{1}{2},-\frac{1}{2},1}\alpha_{\frac{1}{2},-\frac{1}{2},0}
+\sqrt{2}\alpha_{-\frac{1}{2},\frac{1}{2},1}\alpha_{\frac{1}{2},-\frac{1}{2},1}^*
-\sqrt{2}\alpha^*_{-\frac{1}{2},\frac{1}{2},1}\alpha_{\frac{1}{2},-\frac{1}{2},1}\nonumber\\
& -\alpha_{-\frac{1}{2},\frac{1}{2},0}\alpha_{\frac{1}{2},\frac{1}{2},-1}^*
+\alpha^*_{-\frac{1}{2},\frac{1}{2},0}\alpha_{\frac{1}{2},\frac{1}{2},-1}
+\alpha_{\frac{1}{2},-\frac{1}{2},0}\alpha_{\frac{1}{2},\frac{1}{2},-1}^*
-\alpha^*_{\frac{1}{2},-\frac{1}{2},0}\alpha_{\frac{1}{2},\frac{1}{2},-1}\nonumber\\
&\left. -\alpha_{-\frac{1}{2},\frac{1}{2},1}\alpha_{\frac{1}{2},\frac{1}{2},0}^*
+\alpha^*_{-\frac{1}{2},\frac{1}{2},1}\alpha_{\frac{1}{2},\frac{1}{2},0}
+\alpha_{\frac{1}{2},-\frac{1}{2},1}\alpha_{\frac{1}{2},\frac{1}{2},0}^*
-\alpha^*_{\frac{1}{2},-\frac{1}{2},1}\alpha_{\frac{1}{2},\frac{1}{2},0}\right)\nonumber\\
\in & \frac{C_t^{2}}{\sqrt{2}}\times[-2,2]~.
\end{align}
When the separability condition
\begin{align}
    \alpha_{k,j,m}=\beta_k \gamma_j \kappa_m~,~\quad k,j=\pm\frac{1}{2}~,m=-1,0,1~,
\end{align}
is satisfied (indicating no quantum entanglement), the correlation becomes bounded:
\begin{align}
   \left\langle\left(\hat{e}_{W^+}\times\hat{e}_{W^-}\right)\cdot \hat{e}_{e^-}\right\rangle \in  \frac{C_t^{2}}{\sqrt{2}}\times\left[-\tfrac{\sqrt{2}}{2},\tfrac{\sqrt{2}}{2}\right]~.
\end{align}
Therefore, the following condition serves as a sufficient criterion for genuine quantum entanglement in the $t\bar{t}W^-$ system:
\begin{align}
D^\prime_{t\bar{t}W^-}=\frac{\sqrt{2}}{C_t^{2}}\left\langle\left(\hat{e}_{W^+}\times\hat{e}_{W^-}\right)\cdot \hat{e}_{e^-}\right\rangle \in  \left[-2,-\tfrac{\sqrt{2}}{2}\right)\cup\left(\tfrac{\sqrt{2}}{2},2\right]~.
\end{align}

\section{Conclusions} \label{sec:7}

This study establishes a universal framework for detecting QE in collider-produced multi-particle systems, encompassing fermion pairs ($t\bar{t}$, $\tau^{+}\tau^{-}$), bosonic pairs ($W^{-}W^{+}$), and hybrid or three-body systems ($W^{-}t$, $ttt$, $t\bar{t}W^{-}$). By generalizing polarization-state formalisms and exploiting angular correlations in decay products, we derive key observables through phase-space integration and the orthogonality of Wigner $d$-functions, enabling systematic entanglement detection across diverse systems.  

For $t\bar{t}$ systems, the widely used $D = -3\langle\cos\theta_{e^{+}e^{-}}\rangle$ criterion ($D < -\frac{1}{3}$) is shown to be insufficient for detecting QE within the SM at $e^{+}e^{-}$ colliders due to helicity conservation constraints. However, azimuthal correlations, such as $D' =\frac{32}{\pi^2} \langle\cos(\phi_{e^{+}}-\phi_{e^{-}})\rangle$, resolve non-factorizable quantum states, with $D' \in [-1, -\frac{1}{2}) \cup (\frac{1}{2}, 1]$ serving as a sufficient and experimentally accessible criterion for QE. Beyond the SM, scenarios involving heavy Higgs-like scalars ($h'$) or exotic $b'$ quarks enhance entanglement through distinct angular correlations. For instance, $h' \to W^{-}W^{+}$ decays ($m_{h'} < 254.1$ GeV) yield $\langle\cos\theta_{e^{-}e^{+}}\rangle > \frac{1}{4}$, while $b' \to W^{-}t$ decays ($m_{b'} < 319.5$ GeV) produce $D'_{W^{-}t} =\frac{2}{C_t} \langle\cos\theta_{e^{-}W^{+}}\rangle > 1$, directly probing entanglement in these exotic channels.  

For multi-particle systems, such as $ttt$ and $t\bar{t}W^{-}$, we introduce triple-product correlations ($\langle(\hat{e}_1 \times \hat{e}_2) \cdot \hat{e}_3\rangle$) as novel probes of genuine tripartite entanglement. These observables exhibit model-independent sensitivity to quantum correlations, distinguishing entangled states from separable ones and bridging quantum information theory with collider phenomenology.  

Our results provide robust, broadly applicable methodologies for probing QE in high-energy experiments, with immediate applications in SM precision tests and searches for beyond-SM physics. Future work will focus on experimental validation, background suppression strategies, and extending this framework to other systems, to uncover universal entanglement signatures in the quantum realm of particle physics. This approach not only deepens our understanding of quantum phenomena at collider energies but also opens new pathways for discovering exotic particles through their entanglement-driven signatures.

\backmatter

\bmhead{Acknowledgements}

Y. Fang is supported by NSFC Basic Science Centre Program for “Joint Research on High Energy Frontier Particle Physics” (Grant No. 12188102) and National Natural Science Fundation of China under grand No. W2441004. J.~Pei is supported by the National Natural Science Foundation of China under grant No.12247119. L Wu is supported in part by the Natural Science Basic Research Program of Shaanxi, Grant No. 2024JC-YBMS-039 and No. 2024JC-YBMS-521. TL is supported in part by the National Key Research and Development Program of China Grant No. 2020YFC2201504, by the Projects No. 11875062, No. 11947302, No. 12047503, and No. 12275333 supported by the National Natural Science Foundation of China, by the Key Research Program of the Chinese Academy of Sciences, Grant No. XDPB15, by the Scientific Instrument Developing Project of the Chinese Academy of Sciences, Grant No. YJKYYQ20190049, and by the International Partnership Program of Chinese Academy of Sciences for Grand Challenges, Grant No. 112311KYSB20210012. Mustapha Biyabi is supported by the CAS-ANSO Scholarship for Young Talents.

\bibliography{sn-bibliography}

\end{document}